\documentclass[aps,twocolumn,nofootinbib,superscriptaddress]{revtex4}
\usepackage{amsfonts}
\usepackage{amssymb}
\usepackage{amsmath}
\usepackage{graphics}
\usepackage{graphicx}
\usepackage{soul}
\usepackage{natbib}
\usepackage{bm}
\usepackage{mathtools}
\usepackage{stmaryrd}
\usepackage{epstopdf}
\usepackage{color}
\usepackage{lipsum}
\usepackage{balance}
\usepackage{enumitem}
\usepackage[colorlinks=true,citecolor=black,linkcolor=black,urlcolor=black,filecolor=black]{hyperref}

\usepackage{acronym}
\newacro{DF}{distribution function}
\newacroplural{DF}[DFs]{distribution functions}
\newcommand{\DF}{\ac{DF}}
\newcommand{\DFs}{\acp{DF}}
\newacro{BL}{Balescu--Lenard}
\newcommand{\BL}{\ac{BL}}

\newcommand{\rd}{\mathrm{d}}
\newcommand{\re}{\mathrm{e}}
\newcommand{\ri}{\mathrm{i}}
\newcommand{\bw}{\mathbf{w}}
\newcommand{\bwp}{\mathbf{w}^{\prime}}
\newcommand{\bq}{\mathbf{q}}
\newcommand{\bp}{\mathbf{p}}
\newcommand{\bT}{\boldsymbol{\theta}}
\newcommand{\bTp}{\boldsymbol{\theta}^{\prime}}
\newcommand{\bJ}{\mathbf{J}}
\newcommand{\bJp}{\mathbf{J}^{\prime}}
\newcommand{\Mtot}{M_{\mathrm{tot}}}
\newcommand{\Fd}{F_{\mathrm{d}}}
\newcommand{\Hd}{H_{\mathrm{d}}}
\newcommand{\Phid}{\Phi_{\mathrm{d}}}
\newcommand{\deltaD}{\delta_{\mathrm{D}}}
\newcommand{\Uext}{U_{\mathrm{ext}}}
\newcommand{\p}{\partial}
\newcommand{\bk}{\mathbf{k}}
\newcommand{\bkp}{\mathbf{k}^{\prime}}
\newcommand{\psid}{\psi^{\mathrm{d}}}
\newcommand{\bO}{\mathbf{\Omega}}
\newcommand{\bOp}{\mathbf{\Omega}^{\prime}}
\newcommand{\bOpp}{\mathbf{\Omega}^{\prime\prime}}
\newcommand{\tf}{\widetilde{f}}
\newcommand{\mB}{\mathcal{B}}
\newcommand{\ImPart}{\mathrm{Im}}
\newcommand{\tF}{\widetilde{F}}
\newcommand{\tPhi}{\widetilde{\Phi}}
\newcommand{\bare}{\mathrm{bare}}
\newcommand{\dress}{\mathrm{dress}}
\newcommand{\bkpp}{\mathbf{k}^{\prime\prime}}
\newcommand{\bJpp}{\mathbf{J}^{\prime\prime}}
\newcommand{\omegaR}{\omega_{\mathrm{R}}}
\newcommand{\omegaM}{\omega_{\mathrm{M}}}
\newcommand{\mH}{\mathcal{H}}
\newcommand{\tmH}{\widetilde{\mathcal{H}}}
\newcommand{\bx}{\mathbf{x}}
\newcommand{\bv}{\mathbf{v}}
\newcommand{\bvp}{\mathbf{v}^{\prime}}
\newcommand{\tp}{t^{\prime}}
\newcommand{\Res}{\mathrm{Res}}
\newcommand{\bbP}{\mathbb{P}}
\newcommand{\taup}{\tau^{\prime}}
\newcommand{\cst}{\mathrm{cst}}
\newcommand{\mO}{\mathcal{O}}

\newcommand{\mP}{\mathcal{P}}
\newcommand{\bA}{\mathbf{A}}

\newcommand{\rLandau}{\mathrm{Landau}}

\newcommand{\bz}{\mathbf{z}}
\newcommand{\bzp}{\mathbf{z}^{\prime}}
\newcommand{\omegap}{\omega^{\prime}}
\newcommand{\mJ}{\mathcal{J}}
\newcommand{\mJp}{\mathcal{J}^{\prime}}

\usepackage{soul} \usepackage{xcolor}

\begin{document}

\title{Dynamical large deviations for long-range interacting\\inhomogeneous systems without collective effects}

\author{Ouassim Feliachi}
\affiliation{Institut Denis Poisson, Universit\'{e} d’Orl\'{e}ans, CNRS Universit\'{e} de Tours, Orl\'{e}ans, France}
\author{Jean-Baptiste Fouvry}
\affiliation{Institut d'Astrophysique de Paris, UMR 7095, 98 bis Boulevard Arago, F-75014 Paris, France}

\begin{abstract}
We consider the long-term evolution of an inhomogeneous long-range interacting
$N$-body system. Placing ourselves in the dynamically hot limit,
i.e.\ neglecting collective effects,
we derive a large deviation principle for the system's empirical
angle-averaged distribution function.
This result extends the classical ensemble-averaged kinetic theory
given by the so-called inhomogeneous Landau equation,
as it specifies the probability of typical and large dynamical fluctuations.
We detail the main properties of the associated large deviation Hamiltonian,
in particular how it complies with the system's conservation laws
and possesses a gradient structure.
\end{abstract}
\maketitle

\section{Introduction}
\label{sec:Introduction}

As a result of violent relaxation~\citep{LyndenBell1967},
long-range interacting $N$-body systems
generically find themselves to be dynamically frozen
on quasi-stationary states. It is then only as a result
of finite-$N$ fluctuations that these systems can continue
to undergo some statistical relaxation,
driving them closer to their thermodynamical equilibrium.
Such a long-range dynamics covers quite a wide class
of systems like plasmas~\citep{Nicholson1992},
self-gravitating clusters~\citep{Binney+2008},
or even more generic systems~\citep{Campa+2014}.
In the present work, we focus our interest on inhomogeneous systems,
i.e.\ systems with a non-trivial integrable mean-field orbital structure,
as is for example the case in globular clusters~\citep{Heggie+2003}.

In order to describe the long-term evolution of these systems,
a classical starting point is to consider the ensemble-averaged evolution
of the system's \DF\@,
here averaged over a set of initial conditions.
In the context of long-range interacting inhomogeneous systems,
this is described by the inhomogeneous \BL\ equation~\citep{Heyvaerts2010,Chavanis2012}.
In the limit of a dynamically hot system,
i.e.\ a system which only weakly amplifies perturbations,
this kinetic equation reduces to the (simpler) inhomogeneous Landau equation~\citep[see, e.g.\@,][and references therein]{Chavanis2013}.
Both kinetic equations satisfy an $H$-theorem
hence highlighting the irreversibility of the ensemble-averaged dynamics.

Yet, such frameworks, because they solely focus on ensemble-averaged dynamics,
cannot predict the detailed probabilities of typical and large dynamical fluctuations
away from this mean evolution. Such an extension is the realm
of large deviation theory~\citep[see, e.g.\@,][and the detailed review therein]{Bouchet2020}
which describes the statistics of the system's empirical \DF\@.
This is the focus of the present paper.
In particular, we build upon~\cite{Feliachi+2021}
and derive the large deviation Hamiltonian in the case of a dynamically hot
long-range interacting inhomogeneous system.
This calculation therefore generalises the inhomogeneous Landau equation,
which is immediately recovered from the large deviation theory
through an ensemble average.
As the upcoming sections will highlight,
up to a few additional complications stemming from our accounting
of the intricate orbital structure, these calculations
share a lot of similarities with the ones presented in~\cite{Feliachi+2021}.

The paper is organised as follows.
In Sec.~\ref{sec:Dynamics}, we detail our system,
the quasilinear expansion and the inhomogeneous Landau equation.
In Sec.~\ref{sec:LDP},
we derive the system's large deviation Hamiltonian
while neglecting collective effects.
In Sec.~\ref{sec:Properties}, we discuss the main properties
of this Hamiltonian.
We conclude in Sec.~\ref{sec:Conclusion}.
Technical details in the main text are kept to a minimum
and deferred to Appendices.

\section{Dynamics of long-range interacting systems}
\label{sec:Dynamics}

\subsection{System}
\label{sec:System}

We are interested in the long-term evolution
of a long-range interacting Hamiltonian system
in ${2d}$ dimensions.
We denote phase space
with ${ \bw \!=\! (\bq , \bp) }$.
The system is composed of ${ N \!\gg\! 1 }$ particles
of individual mass, ${ m \!=\! \Mtot / N }$,
with $\Mtot$ the system's fixed total active mass.
At any given time, the state of the system can be described
by its empirical \DF\
\begin{equation}
\Fd (\bw , t) = \sum_{i = 1}^{N} m \, \deltaD [\bw \!-\! \bw_{i} (t)] ,
\label{def_Fd}
\end{equation}
with ${ \bw_{i} (t) }$ the location in phase space at time $t$
of particle $i$.
We assume that particles are embedded within some given
external potential, ${ \Uext (\bw) }$
(e.g.\@, the kinetic energy)
and coupled to one another via a long-range pairwise interaction,
${ U (\bw , \bwp) }$.
We denote the typical amplitude of ${ U (\bw , \bwp) }$ with $G$.
The instantaneous specific empirical
Hamiltonian is
\begin{equation}
\Hd \!=\! \Uext (\bw) + \Phid (\bw , t)
\label{def_Hd}
\end{equation}
with the empirical potential
\begin{equation}
\Phid (\bw , t) = \!\! \int \!\! \rd \bwp \, U (\bw , \bwp) \, \Fd (\bwp , t) .
\label{def_Phid}
\end{equation}

The dynamics of $\Fd$ is given by the Klimontovich equation~\citep{Klimontovich1967}.
It reads
\begin{equation}
\frac{\p \Fd}{\p t} + \big[ \Fd , \Hd \big] = 0 ,
\label{Klim_eq}
\end{equation}
with the Poisson bracket
\begin{equation}
\big[ f(\bw) , h (\bw) \big] = \frac{\p f}{\p \bq} \!\cdot\! \frac{\p h}{\p \bp} - \frac{\p f}{\p \bp} \!\cdot\! \frac{\p h}{\p \bq} .
\label{Poisson_bracket}
\end{equation}

In the following, we will consider ensemble averages over the initial conditions
of the $N$ particles. Denoting the ensemble average as ${ \langle \cdot \rangle }$,
we assume then that ${ \langle \Fd (\bw , t) \rangle \!=\! F (\bw , t) }$,
with ${ F (\bw , t) }$ some smooth function,
and ${ H \!=\! \langle \Hd \rangle }$ the associated smooth Hamiltonian.
The mean system
is assumed to be in an integrable stable quasi-stationnary equilibrium.
There exist then canonical angle-action coordinates,
${ (\bT , \bJ) }$~\citep{Binney+2008}, so that
\begin{equation}
F (\bw , t) = F (\bJ , t) ;
\quad
H (\bw , t) = H (\bJ , t) ,
\label{def_AA}
\end{equation}
hence defining the orbital frequencies, ${ \bO (\bJ) \!=\! \p H / \p \bJ }$.
Such a system is said to be in a quasi-stationary equilibrium
since ${ [F (\bJ, t) , H(\bJ, t)] \!=\! 0 }$.

\subsection{Quasilinear expansion}
\label{sec:QuasilinearExpansion}

For a given realisation, we define
\begin{equation}
F_{N} (\bJ , t) = \!\! \int \!\! \frac{\rd \bT}{(2 \pi)^{d}} \, \Fd (\bw , t) ,
\label{def_FN}
\end{equation}
by averaging over the angles.
Following Eq.~\eqref{def_Phid},
we assume that ${ H_{N} \!=\! H_{N} [F_{N}] }$,
only depends on the actions.
Since we have ${ \langle F_{N} \rangle \!=\! F }$,
it is natural to build the decomposition
\begin{subequations}
\begin{align}
\Fd (\bw , t) {} & = F_{N} (\bJ , t) + \frac{1}{\sqrt{N}} \, \delta F (\bw , t) ,
\label{decomposition_QL_F}
\\
\Hd (\bw ,t ) {} & = H_{N} (\bJ , t) + \frac{1}{\sqrt{N}} \, \delta \Phi (\bw , t) ,
\label{decomposition_QL_H}
\end{align}
\label{decomposition_QL}\end{subequations}
where the prefactor ${ 1/\sqrt{N} }$ ensures that the fluctuations,
${ \delta F }$ and ${ \delta \Phi }$, are of order unity w.r.t.\ $N$.

It is crucial to note that Eq.~\eqref{decomposition_QL}
differs from the usual quasilinear decomposition~\citep[see, e.g.\@,][]{Chavanis2012},
${ \Fd \!=\! F \!+\! \delta F / \sqrt{N} }$, performed w.r.t.\
the mean \DF\@, ${ F (\bJ , t) }$. Indeed, in the present approach,
$F_{N}$ remains a stochastic quantity that varies from one realisation to another.
In this work, our goal is to characterise the statistics of the dynamics of ${ F_{N} (\bJ , t) }$
and its deviation away from the mean evolution,
i.e.\ the evolution of ${ F (\bJ , t) }$.

We can now inject Eqs.~\eqref{decomposition_QL}
into Eq.~\eqref{Klim_eq}
to obtain evolution equations for ${ \delta F }$ and ${ F_{N} }$.
We get
\begin{subequations}
\begin{align}
{} & \frac{\p \delta F}{\p \tau} + N \bigg\{ \big[ \delta F , H_{N} \big] + \big[ F_{N} , \delta \Phi \big] \bigg\} = 0 ,
\label{Klim_Fast}
\\
{} & \frac{\p F_{N} (\bJ , \tau)}{\p \tau} + \!\! \int \!\! \frac{\rd \bT}{(2 \pi)^{d}} \, \big[ \delta F , \delta \Phi \big] = 0 ,
\label{Klim_Slow}
\end{align}
\label{Klim_FastSlow}\end{subequations}
where we truncated Eq.~\eqref{Klim_Fast}
at first order in ${ 1/\sqrt{N} }$,
performed an angle average in Eq.~\eqref{Klim_Slow},
and introduced the (slow) time ${ \tau \!=\! t / N }$.
On the one hand, for ${ N \!\gg\! 1 }$,
Eq.~\eqref{Klim_Fast} for ${ \delta F }$
is a fast process, with a timescale for $\tau$ of order ${1/N}$:
it describes the fast dynamics of fluctuations.
On the other hand, Eq.~\eqref{Klim_Slow} is associated with a slow process,
with a timescale for $\tau$ of order $1$: it describes the slow relaxation of orbits.

\subsection{Kinetic equations}
\label{sec:KineticEquations}

In order to describe the ensemble average of Eq.~\eqref{Klim_Slow}
for the asymptotic process of ${ \delta F }$
for fixed $F_{N}$,
one typically proceeds as follows:
(i) Because the angles $\bT$ are ${2\pi}$-periodic,
fluctuations can be decomposed in Fourier space,
hence introducing the associated resonance vectors,
${ \bk , \bkp \!\in\! \mathbb{Z}^{d} }$;
(ii) Assuming a separation between the fast and slow timescales,
one solves Eq.~\eqref{Klim_Fast} for ${ \delta F (\bw , t) }$,
assuming a fixed ${ F_{N} (\bJ , \tau) }$,
see Appendix~\ref{app:LinearDynamics};
(iii) One finally injects these asymptotic expressions
in the r.h.s.\ of Eq.~\eqref{Klim_Slow}, to obtain the kinetic collision operator --
see~\cite{Chavanis2012} for details.

Placing oneself in the dynamically hot limit,
i.e.\ ${ G \!\to\! 0}$,
one finds that the ensemble-averaged long-term evolution
of the system is described
by the inhomogeneous Landau equation~\citep{Chavanis2013} reading
\begin{align}
\frac{\p F (\bJ, \tau)}{\p \tau} = \frac{\p }{\p \bJ} {} & \!\cdot\! \bigg[ \sum_{\bk,\bkp} \bk \!\! \int \!\! \rd \bJp \, B_{\bk\bkp} (\bJ , \bJp)
\label{exp_Landau}
\\
\times {} & \bigg\{ \bk \!\cdot\! \frac{\p F}{\p \bJ} F(\bJp) - \bkp \!\cdot\! \frac{\p F}{\p \bJp} F (\bJ) \bigg\} \bigg] + o (G^{2}),
\nonumber
\end{align}
where we wrote ${ F (\bJ) \!=\! F (\bJ , \tau) }$
to shorten the notations.
We also introduced
\begin{align}
B_{\bk\bkp} (\bJ , \bJp) = {} & \pi (2 \pi)^{d} \Mtot \, \big| \psi_{\bk\bkp} (\bJ , \bJp) \big|^{2}
\label{def_B}
\\
\times {} & \deltaD [\bk \!\cdot\! \bO (\bJ) - \bkp \!\cdot\! \bO (\bJp)] ,
\nonumber
\end{align}
with ${ \deltaD }$ the Dirac delta.
Here, ${ \psi_{\bk\bkp} (\bJ , \bJp , \omega) }$
stands for the bare coupling coefficients
(see Eq.~\ref{def_psi}), which account for the system's inhomogeneity.
When accounting for collective effects,
i.e.\ going beyond the limit ${ G \!\to\! 0 }$,
Eq.~\eqref{exp_Landau} becomes the \BL\ equation~\citep{Heyvaerts2010,Chavanis2012}.
It is obtained from Eq.~\eqref{def_B} by replacing ${ |\psi_{\bk\bkp} (\bJ , \bJp)|^{2} }$
with their dressed analogs, ${ | \psid_{\bk\bkp} (\bJ , \bJp , \bk \!\cdot\! \bO (\bJ)) |^{2} }$,
as defined in Eq.~\eqref{self_psid}.

\section{Large deviation principle}
\label{sec:LDP}

We now want to go beyond the classical computation presented
in Eq.~\eqref{exp_Landau},
by estimating not only the ensemble average of Eq.~\eqref{Klim_Slow},
but the whole cumulant generating function.
This allows one to retrieve not only the average evolution path
for the angle-averaged \DF\@, $F_{N}$,
but the whole probability distribution function
for any evolution path.

Following the same approach as in~\cite{Feliachi+2021},
we can generically estimate the probability that ${ F_{N} (\tau) }$
follows a given time evolution ${ \{ F(\tau) \}_{0 \leq \tau \leq T} }$ through\footnote{In Eq.~\eqref{large_deviations_generic}, we introduce the logarithmic equivalence defined via
${ a_{N} \!\!\!\underset{N \to \infty}{\asymp}\!\!\! \re^{N a} \Longleftrightarrow \!\!\lim\limits_{N \to + \infty}\!\! \ln (a_{N})/N \!=\! a}$.}
\begin{align}
{} & \bbP \big( \big\{ F_{N} (\tau)  \big\}_{0 \leq \tau \leq T} \!=\! \big\{ F (\tau) \big\}_{0 \leq \tau \leq T} \big) \;\; \underset{\mathclap{N \to + \infty}}{\asymp}
\label{large_deviations_generic}
\\
{} & \exp \!\bigg[\! - \frac{N (2 \pi)^{d}}{\Mtot} \sup_{P}  \!\! \int_{0}^{T} \!\!\!\! \rd \tau \, \bigg\{ \bigg(\!\! \int \!\! \rd \bJ  \dot{F} P \bigg) - \mH [F , P] \bigg\} \bigg] ,
\nonumber
\end{align}
with ${ \dot{F} \!=\! \p_{\tau} F }$,
and the prescription that ${ F_{N} (\tau \!=\! 0) }$
converges to ${ F(\tau \!=\! 0) }$ for ${ N \!\to\! + \infty }$.
Equation~\eqref{large_deviations_generic}
also involves the conjugate field, ${ \{ P (\bJ , \tau) \}_{0 \leq \tau \leq T} }$,
over which the maximisation must be performed.
Finally, in Eq.~\eqref{large_deviations_generic},
we introduced the large deviation Hamiltonian
(i.e.\ the scaled cumulant generating function)
\begin{align}
\mH {} & [F , P] = \lim\limits_{\Delta \to + \infty} \frac{\Mtot}{\Delta (2 \pi)^{d}} 
\label{def_mH_generic}
\\
{} & \times \ln \bigg[ \bigg\langle \exp \bigg( \frac{(2 \pi)^{d}}{\Mtot} \!\! \int_{0}^{\Delta} \!\!\!\! \rd t \!\! \int \!\! \rd \bJ \, P (\bJ) \, \p_{\tau} F_{N} [\delta F]  \bigg) \bigg\rangle_{\!F} \bigg] ,
\nonumber
\end{align}
where both ${ F (\bJ) }$ and ${ P (\bJ) }$
are evaluated at time $\tau$,
and ${ \p_{\tau} F_{N} [\delta F] }$ follows from Eq.~\eqref{Klim_Slow}.
Here, ${ \langle \,\cdot\, \rangle_{\!F} }$ denotes an expectation
over the fast process ${ \delta F }$ with ${ F_{N} \!=\! F }$ fixed.
Equations~\eqref{large_deviations_generic} and~\eqref{def_mH_generic}
are generic results describing the large deviations
for the slow evolution of a process driven by a random fast process.
We refer to Appendix~\ref{app:LDP} for a brief heuristic derivation
of Eqs.~\eqref{large_deviations_generic} and~\eqref{def_mH_generic}.

Naturally, here the difficulty lies in the computation
of the average in Eq.~\eqref{def_mH_generic}.
Following the same approach as in~\cite{Feliachi+2021},
we approach this calculation by expanding Eq.~\eqref{def_mH_generic}
w.r.t.\ $G$.
We can write
\begin{equation}
\mH [F , P] = \mH^{(1)} [F , P] + \mH^{(2)} [F, P]  + o (G^{2}).
\label{mH_Landau}
\end{equation}
where ${ \mH^{(1)} }$ (resp.\ ${ \mH^{(2)} }$)
is the first (resp.\ second cumulant).
Computing these cumulants is a cumbersome calculation.
As a first step, this requires the computation
of the asymptotic time evolution of the \DF\ and potential
fluctuations. This is presented in Appendix~\ref{app:LinearDynamics}.
Then, in Appendix~\ref{app:Comp_Cum},
we compute explicitly the various first terms of Eq.~\eqref{mH_Landau}.

The first cumulant reads (Appendix~\ref{app:Cum_1})
\begin{align}
\mH^{(1)} [F , P] = - {} & \sum_{\bk,\bkp} \! \int \!\! \rd \bJ \rd \bJp \,
B_{\bk\bkp} (\bJ , \bJp) \, \bk \!\cdot\! \frac{\p P}{\p \bJ}
\label{mH_1}
\\
\times {} & \bigg[ \bk \!\cdot\! \frac{\p F}{\p \bJ} F(\bJp) - \bkp \!\cdot\! \frac{\p F}{\p \bJp} F (\bJ) \bigg]   + o (G^{2}) ,
\nonumber
\end{align}
while the second cumulant reads
(Appendix~\ref{app:Cum_2})
\begin{align}
\mH^{(2)} [F , P] = {} & \sum_{\bk,\bkp} \! \int \!\! \rd \bJ \rd \bJp \, B_{\bk\bkp} (\bJ , \bJp) \, \bk \!\cdot\! \frac{\p P}{\p \bJ} 
\label{mH_2}
\\
\times {} & \bigg[ \bk \!\cdot\! \frac{\p P}{\p \bJ} - \bkp \!\cdot\! \frac{\p P}{\p \bJp} \bigg] F (\bJ) \, F (\bJp) + o (G^{2}) .
\nonumber
\end{align}

Finally, in Appendix~\ref{app:Cum_next}, we justify why all cumulants
beyond the two first ones can be neglected in the dynamically hot limit,
i.e.\ they are all of order ${ o (G^{2}) }$.

Equation~\eqref{large_deviations_generic}
in conjunction with Eq.~\eqref{mH_Landau} is the main result of this section.
It characterises the dynamical large deviations
of the angle-averaged \DF\@, ${ F_{N} (\bJ , \tau) }$.
We discuss their main properties in Sec.~\ref{sec:Properties}.

Unfortunately, the expansion from Eq.~\eqref{mH_Landau}
stops being effective when accounting for collective effects,
i.e.\ in the dynamically cold limit. In that case, one must resort to an explicit
computation of the exponential average from Eq.~\eqref{def_mH_generic},
as all cumulants in Eq.~\eqref{def_mH_generic} contribute to the large deviation Hamiltonian.
\cite{Feliachi+2022} succeeded in performing this calculation
in the case of a homogeneous system.
Unfortunately, the theorems and methods
used therein do not lend themselves straightforwardly
to the inhomogeneous case.
One reason of these additional difficulties lies in the fact 
that in homogeneous systems, the resonance condition,
${ \deltaD (\bk \!\cdot\! [\bv \!-\! \bvp]) }$, is diagonal
w.r.t.\ the resonance number $\bk$. This is not the case
anymore in inhomogeneous systems where the resonance condition
becomes ${ \deltaD [\bk \!\cdot\! \bO (\bJ) \!-\! \bkp \!\cdot\! \bO (\bJp)] }$,
as in Eq.~\eqref{def_B}.
This will be the topic of a future work.

\section{Properties}
\label{sec:Properties}

In this section, we briefly present the main properties
of the large deviation Hamiltonian from Eq.~\eqref{mH_Landau}.
We refer to Appendix~\ref{app:Properties} for technical details
and to sec.\@~{2.1.1} of~\cite{Feliachi+2021}
for thorough discussions.

\subsection{Most probable path}
\label{sec:MostProbable}

As expected, the most probable path
\begin{equation}
\frac{\p F (\bJ , \tau)}{\p \tau} = \frac{\delta \mH}{\delta P (\bJ)} [F , P \!=\! 0] ,
\label{most_probable}
\end{equation}
yields exactly the inhomogeneous Landau equation~\eqref{exp_Landau}.
Equation~\eqref{most_probable} is the Hamilton equation prescribing the evolution path
that minimizes the large deviation action,
i.e.\ the most probable evolution path for the empirical \DF\@~\citep[see sec.\@~{7.2.2} in][]{Bouchet2020}.

\subsection{Conservation laws}
\label{sec:Conservation}

If ${ C[F] }$ is a conserved quantity for the $N$-body dynamics,
the large deviation Hamiltonian must satisfy
\begin{equation}
\!\! \int \!\! \rd \bJ \, \frac{\delta C[F]}{\delta F(\bJ)} \, \frac{\delta \mH}{\delta P (\bJ)} = 0 .
\label{cons_C}
\end{equation}
This symmetry implies that the large deviation action
(i.e.\ the exponential argument in the r.h.s.\ of Eq.~\ref{large_deviations_generic})
is infinite for evolution paths
that do not satisfy ${ \! \int \! \rd \bJ \dot{F} \, \delta C[F] / \delta F(\bJ) \!=\! 0 }$.
Phrased differently, any evolution path for the \DF\
violating the conservation laws has zero probability~\citep[see sec.\@~{7.2.6} in][]{Bouchet2020}.
In Appendix~\ref{app:Conservation}, we explicitly show that
Eq.~\eqref{mH_Landau} is consistent with the mass
and energy conservation, defined via
\begin{subequations}
\begin{align}
M [F] {} & = \!\! \int \!\! \rd \bJ \, F (\bJ)  
 {} & \text{(mass conservation)},
\label{def_C_mass}
\\
E [F] {} & = \!\! \int \!\! \rd \bJ \, H (\bJ) \, F (\bJ) 
 {} & \text{(energy conservation)} .
\label{def_C_energy}
\end{align}
\label{def_C}\end{subequations}

\subsection{Hamilton--Jacobi equation}
\label{sec:HamiltonJacobi}

We define the system's entropy as
\begin{equation}
S [F] = - \!\! \int \!\! \rd \bJ \, F (\bJ) \, \ln [F (\bJ)] .
\label{def_S}
\end{equation}
As detailed in Appendix~\ref{app:HamiltonJacobi}, one can show that
it solves the stationary Hamilton--Jacobi equation
\begin{equation}
\mH \big[ F , - \delta S / \delta F \big] = 0 ,
\label{HJ_equation}
\end{equation}
where ${ \delta S / \delta F }$ stands for the function
${ J \!\mapsto\! \delta S[F] / \delta F (\bJ) }$.
Yet, Eq.~\eqref{HJ_equation} is not sufficient
to prove that the negative of the entropy is the quasipotential associated
with the present large deviation principle~\citep[see sec.\@~{7.2.3} in][]{Bouchet2020}.
Indeed, one also has to prove that ${ S[F] }$
has an unique maximum:
this is typically not true
for self-gravitating systems~\citep[see, e.g.\@,][]{Padmanabhan1990,Chavanis2006}.

\subsection{Time-reversal symmetry}
\label{sec:TimeReversal}

The large deviation Hamiltonian complies
with the generalised time-reversal symmetry
(see Appendix~\ref{app:TimeReversal})
\begin{equation}
\mH [F , - P] = \mH [F , P - \delta S / \delta F] .
\label{time_reversal}
\end{equation}
Such a relation corresponds to a detailed balance condition
at the level of large deviations,
associated with the symmetry ${ \tau \!\to\! T \!-\! \tau }$
in Eq.~\eqref{large_deviations_generic}~\citep[see sec.\@~{7.3.1} in][]{Bouchet2020}.

\subsection{Gradient structure}
\label{sec:Gradient}

The large deviation Hamiltonian generically induces a gradient flow~\cite{Mielke+2014}.
Indeed, as detailed in Appendix~\ref{app:Gradient},
the second cumulant from Eq.~\eqref{mH_2}
can be written as
\begin{equation}
\mH^{(2)} [F , P] = \!\! \int \!\! \rd \bJ \rd \bJp \, P (\bJ) \, P (\bJp) \, Q [F] (\bJ , \bJp) ,
\label{H2_from_Q}
\end{equation}
where ${ Q [F] }$ reads
\begin{align}
Q {} & [F] (\bJ , \bJp) \!=\! \sum_{\bk , \bkp} \bk \!\cdot\! \frac{\p }{\p \bJ} \bigg\{ \frac{\p }{\p \bJp} \!\cdot\! \!\bigg[\! - \!\bkp F (\bJ) F (\bJp) B_{\bk\bkp} \!(\bJ , \bJp)
\nonumber
\\
{} & +  \bk \, \deltaD (\bJ \!-\! \bJp) F (\bJ) \!\! \int \!\! \rd \bJpp F (\bJpp) B_{\bk\bkp} (\bJ , \bJpp) \bigg] \bigg\} .
\label{def_Q}
\end{align}

The inhomogeneous Landau Eq.~\eqref{exp_Landau}
can then be rewritten as
\begin{equation}
\frac{\p F (\bJ , \tau)}{\p \tau} = \!\! \int \!\! \rd \bJp \, Q [F] (\bJ , \bJp) \, \frac{\delta S[F]}{\delta F (\bJp)} ,
\label{Grad_Landau}
\end{equation}
with the entropy, ${ S[F] }$, defined in Eq.~\eqref{def_S}.
Given that $H[F,P]$ is convex w.r.t.\ $P$,
$Q[F]$ is a positive operator on the space of \DFs\@.
Therefore, Eq.~\eqref{Grad_Landau}
illustrates the increase of entropy along solutions
of the Landau equation~\citep[see sec.\@~{5} in][]{Bouchet2020}.

\subsection{Stochastic Landau equation}
\label{sec:StochasticLandau}

We note that the large deviation Hamiltonian from Eq.~\eqref{mH_Landau}
is quadratic in ${ P (\bJ , t) }$. This implies that large deviations
are Gaussian~\citep[see, e.g.\@, sec.\@~{4.2} in][]{Feliachi+2021}.
As a result, as detailed in Appendix~\ref{app:Stochastic},
one can setup a stochastic partial differential equation
which obeys the large deviation principle from Eq.~\eqref{large_deviations_generic}.
It reads
\begin{align}
{} & \frac{\p F_{N} (\bJ , \tau)}{\p \tau} = \bigg[\! \frac{\p F_{N} (\bJ , \tau) }{\p \tau} \!\bigg]_{\rLandau} + \zeta[F_{N}] (\bJ , \tau) ,
\label{SPDE}
\end{align}
where ${ [\p F_{N} (\bJ) / \p \tau ]_{\rLandau} }$
is the inhomogeneous Landau collision operator,
i.e.\ the r.h.s.\ of Eq.~\eqref{exp_Landau}.
We also introduced the Gaussian random field, ${ \zeta [F] (\bJ , \tau) }$.
Following sec.~{4.2} in~\cite{Feliachi+2021},
it obeys
\begin{subequations}
\begin{align}
\big\langle \zeta [F] (\bJ , \tau) \big\rangle {} & = 0 ,
\label{def_zeta_1}
\\
\big\langle \zeta [F] (\bJ , \tau) \, \zeta [F] (\bJp , \taup) \big\rangle {} & \!=\! \frac{2 m}{(2 \pi)^{d}} \, Q [F] (\bJ , \bJp) \, \deltaD (\tau \!-\! \taup) ,
\label{def_zeta_2}
\end{align}
\label{def_zeta}\end{subequations}
where averages are taken at fixed $F$.

Although intricate, Eq.~\eqref{SPDE} is an important result
as it allows one to ``mimic'' directly the stochastic evolution
of $F_{N}$ and its large deviations,
without ever integrating the $N$-body equations of motion.
Following Sec.~\ref{sec:Conservation},
Eq.~\eqref{SPDE} exactly conserves
the total mass and total energy.
Finally, in Appendix~\ref{app:Stochastic} (see Eq.~\ref{diag_zeta}),
we present an alternative writing of ${ \zeta [F] (\bJ , \tau) }$ which
(i) eases the effective sampling of the stochastic noise,
(ii) highlights explicitly the compliance with the conservation laws.

\subsection{Homogeneous limit}
\label{sec:Homogeneous}

It is straightforward to recover the results from~\cite{Feliachi+2021}
in the limit of a (multi-periodic) homogeneous system.
In that case, (i) the angle-action coordinates become ${ (\bT,\bJ) \!\to\! (\bx , \bv) }$,
the volume elements ${ (2\pi)^{d} \!\to\! L^{3} }$,
with $L$ the size of the box,
and ${ N / \Mtot }$ plays the role of the plasma parameter, $\Lambda$;
(ii) the bare coupling coefficients are constrained by symmetry
and independent of the orbits, namely
 ${ \psi_{\bk\bkp} (\bJ , \bJp) \!\to\! \delta_{\bk\bkp} / |\bk|^{2} }$,
 up to a prefactor.
Equation~\eqref{mH_Landau} then falls back
on the result from~\cite{Feliachi+2021}.

\section{Conclusion}
\label{sec:Conclusion}

In this paper, we investigated dynamical large deviations
in systems with long-range interactions.
In the view of generalising~\cite{Feliachi+2021}, we focused here
on the case of inhomogeneous systems,
i.e.\ systems with a non-trivial orbital structure.
Starting from the generic large deviation Hamiltonian
of a slow-fast system
and placing ourselves in the dynamically hot limit,
i.e.\ neglecting collective effects,
we computed the two first cumulants
of the associated large deviation Hamiltonian.
Equation~\eqref{mH_Landau} is the main result of this work,
and encodes the likelihood
of any given evolution path for the system's
angle-averaged \DF\@.
In Sec.~\ref{sec:Properties}, we highlighted that Eq.~\eqref{mH_Landau}
complies with all the expected properties,
such as (i) recovering the inhomogeneous Landau equation~\citep{Chavanis2013}
in the ensemble-averaged limit;
(ii) ensuring the conservation laws;
(iii) and possessing a natural gradient structure.
Finally, we emphasised that the quadratic dependence
of the large deviation Hamiltonian
w.r.t.\ the conjugate field allows one
to construct an effective stochastic partial differential equation
with the expected dynamical large deviations.

The present work is only one step toward
ever more detailed description of the statistical
properties of dynamical large deviations
in long-range interacting inhomogeneous systems.
We conclude by mentioning a few possible venues
for future explorations.

Naturally, it would be useful
to generalise Eq.~\eqref{mH_Landau}
and lift the assumption ${ G \!\to\! 0 }$,
so as to generalise~\cite{Feliachi+2022}
to the inhomogeneous case.
Unfortunately, this is no easy undertaking as both the resonance condition
and the coupling coefficients get more intricate
as one considers an inhomogeneous system.
Similarly, approaching these calculations through
the BBGKY hierarchy~\citep[see, e.g.\@,][]{Nicholson1992}
could also prove insightful.

The generic probability distribution from Eq.~\eqref{large_deviations_generic}
is challenging to evaluate in practice. Though, it would be enlightening
to compute it numerically, starting with some simple long-range interacting systems.
One could begin with the Hamiltonian Mean Field model~\citep{Antoni+1995}
in an inhomogeneous configuration,
for example tracking the statistics of the large deviations
of the system's magnetisation.
Similarly, one could extend the present work
to long-range interacting systems submitted to a stochastic forcing~\citep[see, e.g.\@,][]{Nardini+2012}.
These systems exhibit phase transitions
whose most likely transition path
could be recovered via a large deviation principle --
through the resolution of a system of coupled partial
differential equations for ${ F(\bJ , \tau) }$ and ${ P (\bJ , \tau) }$.
Such preliminary explorations
are mandatory first steps before applying the present statistical approach
to more realistic systems.

Here, the system was always assumed
to be (strongly) linearly stable.
This allowed us to neglect any possible contributions from the system's
damped modes in the linear dynamics of fluctuations
(see Appendix~\ref{app:TimeEvolution}). In the case of weakly stable systems,
it would be interesting to extend the present calculation
to account for the effects of slowly decaying modes and
wave-particle interactions~\citep[see, e.g.\@,][]{Hamilton+2020},
as well as possible dynamical phase transitions
toward linear instability~\citep[see, e.g.\@,][]{DeRijcke+2019}.
Additionally, one could investigate
the connections between the present dynamical large deviations
and the thermal (van--Kampen) fluctuations considered
in~\cite{Lau+2021_I,Lau+2021_II}.

Finally, we limited ourselves to diffusion sourced by two-body correlations,
i.e.\ ${1/N}$ effects.
Yet, in the case of one-dimensional systems,
such a relaxation can identically vanish.
This is a kinetic blocking~\citep[see, e.g.\@,][]{Bouchet+2005}
and these systems can only relax via the weaker
${ 1/N^{2} }$ effects, i.e.\ three-body correlations~\citep[see, e.g.\@,][]{Fouvry2022}.
It would be informative to investigate the properties
of dynamical large deviations in such contrived systems.

\begin{acknowledgments}
This work is partially supported by the grant Segal ANR-19-CE31-0017 of the French Agence Nationale de la Recherche and by the Idex Sorbonne Universit\'{e}.
We are grateful to J.\ Barr\'{e}, C.\ Pichon, M.\ Roule for insightful discussions.
\end{acknowledgments}

\appendix

\section{Linear dynamics of fluctuations}
\label{app:LinearDynamics}

In this Appendix,
we solve for the linear dynamics of the fluctuations,
${ \delta F }$, as driven by Eq.~\eqref{Klim_Fast}.
The calculations below follow closely~\cite{Chavanis2012,Fouvry+2018}.
In this Appendix, the angle-averaged \DF\@, $F_{N}$,
has simply been denoted with, $F$,
to shorten the notations.

\subsection{Laplace--Fourier transforms}
\label{app:LaplaceFourier}

We define the Fourier transform w.r.t.\ the angle $\bT$ via
\begin{subequations}
\begin{align}
f (\bw , t) {} & = \sum_{\mathclap{\bk \in \mathbb{Z}^{d}}} f_{\bk} (\bJ , t) \, \re^{\ri \bk \cdot \bT} ,
\label{def_Fourier_back}
\\
f_{\bk} (\bJ , t) {} & = \!\! \int \!\! \frac{\rd \bT}{(2 \pi)^{d}} f (\bw , t) \, \re^{- \ri \bk \cdot \bT} .
\label{def_Fourier_frwd}
\end{align}
\label{def_Fourier}\end{subequations}
The self-consistency relation from Eq.~\eqref{def_Phid}
when used for the fluctuations then reads
\begin{equation}
\delta \Phi_{\bk} (\bJ , t) = (2 \pi)^{d} \sum_{\bkp} \!\! \int \!\! \rd \bJp \, \psi_{\bk\bkp} (\bJ , \bJp) \, \delta F_{\bkp} (\bJp , t) .
\label{intro_psikk}
\end{equation}
In that expression, ${ \psi_{\bk\bkp} (\bJ , \bJp) }$ are the bare coupling coefficients.
They follow from the expansion of the pairwise interaction potential as
\begin{equation}
U (\bw , \bwp) = \sum_{\bk , \bkp} \psi_{\bk\bkp} (\bJ , \bJp) \, \re^{\ri (\bk \cdot \bT - \bkp \cdot \bTp)} ,
\label{def_psi}
\end{equation}
These coefficients satisfy the two symmetries
\begin{subequations}
\begin{align}
\psi_{\bkp\bk} (\bJp , \bJ) {} & = \psi_{\bk\bkp}^{*} (\bJ , \bJp) ,
\label{sym_psi_order}
\\
\psi_{-\bk-\bkp} (\bJ , \bJp) {} & = \psi_{\bk\bkp}^{*} (\bJ , \bJp) .
\label{sym_psi_sign}
\end{align}
\label{symmetry_psi}\end{subequations}

We define the Laplace transform with the convention
\begin{equation}
\tf (\omega) = \!\! \int_{0}^{+ \infty} \!\!\!\! \rd t \, f (t) \, \re^{\ri \omega t} ;
\quad
f (t) = \!\! \int_{\mB} \!\! \frac{\rd \omega}{2 \pi} \, \tf (\omega) \, \re^{- \ri \omega t} ,
\label{def_Laplace}
\end{equation}
where the Bromwich contour, $\mB$,
has to pass above all the poles of the integrand,
i.e.\ ${ \ImPart[\omega] }$ has to be large enough.
When expressed in Fourier--Laplace space, Eq.~\eqref{Klim_Fast} becomes
\begin{equation}
\delta \tF_{\bk} (\bJ , \omega) = - \frac{\bk \!\cdot\! \p F / \p \bJ}{\omega \!-\! \bk \!\cdot\! \bO} \, \delta \tPhi_{\bk} (\bJ , \omega) - \frac{\delta F_{\bk} (\bJ , 0)}{\ri (\omega \!-\! \bk \!\cdot\! \bO)} 
\label{Klim_Fast_FL}
\end{equation}
with ${ \bO \!=\! \bO(\bJ) }$
and ${ \delta F_{\bk} (\bJ , 0) }$ describing the fluctuations
in the \DF\ at the initial time.
Once again, we recall that for the sake of shorter notations,
we wrote the angle-averaged \DF\@, $F_{N}$, as $F$. 

\subsection{Self-consistency}
\label{app:SelfConsistency}

We now act on both sides of Eq.~\eqref{Klim_Fast_FL}
with the same operator as in the r.h.s.\ of Eq.~\eqref{intro_psikk}. We get
\begin{align}
{} & \delta \tPhi_{\bk} (\bJ , \omega) \!=\! - (2 \pi)^{d} \! \sum_{\bkp} \!\! \int \!\!\! \rd \bJp \!\frac{\psi_{\bk\bkp} (\bJ , \bJp)}{\omega \!-\! \bkp \!\!\cdot\! \bOp} \bkp \!\!\cdot\! \frac{\p F}{\p \bJp} \delta \tPhi_{\bkp} \!(\bJp \!, \omega)
\nonumber
\\
{} & \quad\quad\quad - (2 \pi)^{d} \! \sum_{\bkp} \!\! \int \!\! \rd \bJp \frac{\psi_{\bk \bkp} (\bJ , \bJp) }{\ri (\omega \!-\! \bkp \!\cdot\! \bOp)} \delta F_{\bkp} (\bJp , 0) ,
\label{self_deltaPhi}
\end{align}
with ${ \bOp \!=\! \bO (\bJp) }$.

In the absence of collective effects
(i.e.\ ${ G \!\to\! 0 }$),
the first term in the r.h.s.\ of Eq.~\eqref{self_deltaPhi}
can be neglected, to get
\begin{equation}
\delta \tPhi_{\bk}^{\bare} (\bJ , \omega) \!=\! - (2 \pi)^{d} \sum_{\bkp} \!\! \int \!\! \rd \bJp \, \frac{\delta F_{\bkp} (\bJp , 0) \, \psi_{\bk\bkp} (\bJ , \bJp)}{\ri (\omega \!-\! \bkp \!\cdot\! \bOp)} .
\label{delta_Phi_bare}
\end{equation}
When collective effects are accounted for,
we may assume that the dressed potential perturbations
follow the ansatz
\begin{equation}
\delta \tPhi_{\bk}^{\dress} (\bJ , \omega) \!=\! - (2 \pi)^{d} \sum_{\bkp} \!\! \int \!\! \rd \bJp \, \frac{\delta F_{\bkp} (\bJp , 0) \, \psid_{\bk\bkp} (\bJ , \bJp , \omega)}{\ri (\omega \!-\! \bkp \!\cdot\! \bOp)} ,
\label{delta_Phi_dressed}
\end{equation}
where the frequency-dependent dressed coupling coefficients,
${ \psid_{\bk\bkp} (\bJ , \bJp , \omega) }$, remain to be determined.
When injected into Eq.~\eqref{self_deltaPhi},
we find that the dressed coupling coefficients satisfy
the self-consistent relation
\begin{align}
{} & \psid_{\bk\bkp} (\bJ , \bJp , \omega) = \psi_{\bk\bkp} (\bJ , \bJp)
\label{self_psid}
\\
- {} & (2 \pi)^{d} \sum_{\bkpp} \!\! \int \!\! \rd \bJpp \frac{\psi_{\bk\bkpp} (\bJ , \bJpp) \, \bkpp \!\!\cdot\! \p F / \p \bJpp }{\omega \!-\! \bkpp \!\cdot\! \bOpp} \psid_{\bkpp\bkp} (\bJpp , \bJp , \omega) ,
\nonumber
\end{align}
with ${ \bOpp \!=\! \bO (\bJpp) }$.
Since Eq.~\eqref{Klim_Fast_FL} was derived assuming ${ \ImPart[\omega] \!>\! 0 }$ large enough,
for ${ \omegaR \!\in\! \mathbb{R} }$, the resonant denominator
in Eq.~\eqref{self_psid} has to be interpreted as
\begin{equation}
\frac{1}{\omegaR \!-\! \bkpp \!\cdot\! \bOpp} \to \frac{1}{\omegaR \!-\! \bkpp \!\cdot\! \bOpp \!+\! \ri \gamma} ,
\label{regularisation_denominator}
\end{equation}
with ${ \gamma \!\to\! 0^{+} }$.
We refer to~\cite{Fouvry+2022}
(and references therein) for a discussion
of the associated Landau's prescription.
We note that an explicit expression for $\psid_{\bk\bkp}$
can be obtained using the basis method~\citep[see, e.g.\@,][and references therein]{Fouvry+2018},
but this will not needed here.

Ultimately, we find that the coupling coefficients asymptotically scale
w.r.t.\ $G$, the amplitude of the pairwise interaction, like
\begin{subequations}
\begin{align}
\psi_{\bk\bkp} {} &\propto G ,
\label{scaling_psi}
\\
\psid_{\bk\bkp} {} & \propto \frac{\psi_{\bk\bkp}}{1 \!-\! \psi_{\bk\bkp}} \propto \frac{G}{1 \!-\! G} .
\label{scaling_psid}
\end{align}
\label{scaling_psi_psid}\end{subequations}

\subsection{Time evolution}
\label{app:TimeEvolution}

Taking the inverse Laplace transform of Eq.~\eqref{delta_Phi_dressed},
the dressed fluctuations generically evolve according to
\begin{align}
\delta \Phi_{\bk} (\bJ , t) = - {} & (2 \pi)^{d} \sum_{\bkp} \!\! \int \!\! \rd \bJp \, \delta F_{\bkp} (\bJp , 0)
\nonumber
\\
\times {} & \int_{\mB} \!\! \frac{\rd \omega}{2 \pi} \frac{\psid_{\bk\bkp} (\bJ , \bJp , \omega)}{\ri (\omega \!-\! \bkp \!\cdot\! \bOp)} \, \re^{- \ri \omega t} ,
\label{inv_deltaPhi}
\end{align}
where the Bromwich contour, $\mB$,
has to pass above all the poles of the integrand
in the complex $\omega$-plane.
Because we assumed that the system is linearly stable,
the function ${ \omega \!\mapsto\! \psid_{\bk\bkp} (\bJ , \bJp , \omega) }$
only has poles in the lower half of the complex plane.
These are associated with Landau damped modes
and correspond to frequencies $\omegaM$
with ${ \ImPart[\omegaM] \!<\! 0 }$.

As usual, we proceed by distorting the contour $\mB$
to the lower half of the complex plane so that ${ |\re^{- \ri \omega t}| \!\to\! 0 }$,
snagging onto the poles of the integrand.
In Eq.~\eqref{inv_deltaPhi}, there is a single pole along the real axis,
namely in ${ \omega \!=\! \bkp\!\cdot\! \bOp }$.
Paying attention to the direction of integration, each pole contributes
a ${ - 2 \ri \pi \Res[...] }$.
Placing ourselves in the limit ${ t \!\gg\! |1 / \ImPart[\omegaM]| }$,
we neglect the contributions from the damped modes~\citep[see, e.g.\@,][]{Hamilton+2020}.
Once these have faded away,
Eq.~\eqref{inv_deltaPhi} becomes
\begin{align}
\delta \Phi_{\bk} (\bJ , t) = {} & (2 \pi)^{d} \sum_{\bkp} \!\! \int \!\! \rd \bJp \, \re^{- \ri \bkp \cdot \bOp t} \, \delta F_{\bkp} (\bJp , 0)
\nonumber
\\
{} & \times \psid_{\bk\bkp} (\bJ , \bJp , \bkp \!\cdot\! \bOp) .
\label{time_deltaPhi}
\end{align}

Having determined the time-evolution of the potential fluctuations,
we now set out to determine the time-evolution of the \DF\ fluctuations themselves.
An efficient approach to perform this calculation
is to rely on the self-consistency relation from Eq.~\eqref{intro_psikk}.
We start with Eq.~\eqref{time_deltaPhi} in which we replace
${ \psid_{\bk\bkp} }$ with the r.h.s.\ from Eq.~\eqref{self_psid}.
We get
\begin{align}
{} & \delta \Phi_{\bk} (\bJ , t) \!=\! (2 \pi)^{d} \sum_{\bkp} \!\! \int \!\! \rd \bJp \re^{- \ri \bkp \cdot \bOp t} \delta F_{\bkp} (\bJp , 0) \, \psi_{\bk\bkp} (\bJ , \bJp)
\nonumber
\\
{} & - (2 \pi)^{2d} \sum_{\bkp , \bkpp} \!\! \int \!\! \rd \bJp \rd \bJpp \re^{- \ri \bkp \cdot \bOp t} \delta F_{\bkp} (\bJp , 0) \, \psi_{\bk\bkpp} (\bJ , \bJpp)
\nonumber
\\
{} & \times \frac{\psid_{\bkpp \bkp} (\bJpp , \bJp , \bkp \!\cdot\! \bOp)}{\bkp \!\cdot\! \bOp \!-\! \bkpp \!\cdot\! \bOpp \!+\! \ri \gamma} \, \bkpp \!\cdot\! \frac{\p F}{\p \bJpp}  .
\label{calc_time_dF}
\end{align}
We now perform the switch ${ (\bkp , \bJp) \!\leftrightarrow\! (\bkpp , \bJpp) }$
in the second term. Factorising the integrand with ${ \psi_{\bk\bkp} (\bJ , \bJp) }$,
we finally rely on Eq.~\eqref{intro_psikk}
to identify the remainder of the integrand with ${ \delta F_{\bkp} (\bJp , t) }$.
Ultimately, we get
\begin{align}
\delta F_{\bk} (\bJ , t) = {} & \re^{- \ri \bk \cdot \bO t} \, \delta F_{\bk} (\bJ , 0)
\label{time_dF}
\\
+ {} & (2 \pi)^{d} \bk \!\cdot\! \frac{\p F}{\p \bJ} \sum_{\bkp} \!\! \int \!\! \rd \bJp \, \re^{- \ri \bkp \cdot \bOp t} \, \delta F_{\bkp} (\bJp , 0)
\nonumber
\\
\times {} & \frac{\psid_{\bk\bkp} (\bJ , \bJp , \bkp \!\cdot\! \bOp)}{\bk \!\cdot\! \bO \!-\! \bkp \!\cdot\! \bOp \!-\! \ri \gamma} ,
\nonumber
\end{align}
with ${ \gamma \!\to\! 0^{+} }$.

Glancing back at Eq.~\eqref{scaling_psi_psid},
we find from Eqs.~\eqref{time_deltaPhi} and~\eqref{time_dF}
that the potential and \DF\ fluctuations scale asymptotically w.r.t.\ $G$ like
\begin{subequations}
\begin{align}
\delta \Phi (t) {} & \propto \delta F (0) \, \frac{G}{1 \!-\! G} ,
\label{scaling_deltaPhi}
\\
\delta F (t) {} & \propto \delta F (0) + \delta F (0) \, \frac{G}{1 \!-\! G} .
\label{scaling_deltaF}
\end{align}
\label{scaling_deltaPhi_deltaF}\end{subequations}
These scalings play an important role in easing the computation
of the cumulants, as detailed in Appendix~\ref{app:Comp_Cum}.

\section{Large deviation principle}
\label{app:LDP}

In this Appendix, we present a brief (very) heuristic derivation
of the generic large deviation principle from Eqs.~\eqref{large_deviations_generic} and~\eqref{def_mH_generic}.
We refer to sec.\@~{2.2} in~\cite{Feliachi+2021} and references therein
for a much more thorough presentation.

\subsection{G\"{a}rtner--Ellis theorem}
\label{app:GartnerEllis}

This section is inspired from sec.\@~{3.3} of~\cite{Touchette2009}.
As first step, let us mimic the empirical \DF\ from Eq.~\eqref{def_Fd}
and consider a real random variable, $F_{N}$,
parametrised by $N$.
From it, we define the scaled cumulant generating function
\begin{equation}
\mH [P] = \frac{1}{N} \ln \big[ \big\langle \re^{N F_{N} P} \big\rangle \big] ,
\label{calc_Gartner_II}
\end{equation}
also called the large deviation Hamiltonian,
which is assumed to be finite.

In the limit ${ N \!\gg\! 1 }$, we assume that ${ F_{N} }$
follows a large deviation principle of the form
\begin{equation}
\bbP (F_{N} \!=\! F) \;\;\;\underset{\mathclap{N \to + \infty}}{\asymp} \;\;\; \re^{- N I [F]} , 
\label{LDP_Gartner}
\end{equation}
where $N$ is the large deviation rate
and ${ I [F] }$ the large deviation function.
For a given $P$, we can compute
\begin{align}
\big\langle \re^{N F_{N} P} \big\rangle {} & \;\;\; = \!\! \int \!\! \rd F \, \re^{N F P} \, \bbP (F_{N} \!=\! F)
\nonumber
\\
{} & \;\;\;\underset{\mathclap{N \to + \infty}}{\asymp} \;\;\; \int \!\! \rd F \, \re^{N (F P - I [F])} .
\label{calc_Gartner_Ia}
\end{align}
Since ${ N \!\gg\! 1 }$, we estimate this integral using the saddle-point method.
The dominating contribution comes from the maximum
of the argument of the exponential to give
\begin{equation}
\big\langle \re^{N F_{N} P} \big\rangle \;\;\;\underset{\mathclap{N \to + \infty}}{\asymp} \;\;\; 
\exp \!\big[ N \sup_{F} \big\{ F P - I [F] \big\} \big] .
\label{calc_Gartner_Ib}
\end{equation}

Comparing Eqs.~\eqref{calc_Gartner_Ib} and~\eqref{calc_Gartner_II},
we asymptotically obtain the relation ${ \mH [P] \!=\! \sup_{F} \{ F P \!-\! I [F] \} }$.
For a differentiable ${ H [P] }$,
this Legendre transform is involutive,
and we finally obtain the large deviation function as
\begin{equation}
I [F] = \sup_{P} \big\{ F P - \mH [P] \big\} .
\label{Legendre_2}
\end{equation}
This is the G\"artner--Ellis theorem~\citep[see, e.g.\@,][ and references therein]{Touchette2009},
and constitutes the foundation on which Eqs.~\eqref{large_deviations_generic}
and~\eqref{def_mH_generic} lie.

\subsection{Slow-fast systems}
\label{app:SlowFast}

Let us now mimic the evolution equations at play,
namely Eqs.~\eqref{Klim_FastSlow},
by considering two coupled stochastic variables,
${ (F_{N} , \delta F) }$, of order unity and evolving according to
\begin{subequations}
\begin{align}
\frac{\p \delta F}{\p \tau} {} & = N \, \frac{\p \delta F}{\p \tau} [F_{N} , \delta F]
\label{Klim_Fast_LDP}
\\
\frac{\p F_{N}}{\p \tau} {} & = \frac{\p F_{N}}{\p \tau} [F_{N} , \delta F] ,
\label{Klim_Slow_LDP}
\end{align}
\label{Klim_FastSlow_LDP}\end{subequations}
with the slow time, ${ \tau \!=\! t / N }$.
Owing to the presence of the factor ${ N }$ in Eq.~\eqref{Klim_Fast_LDP},
this constitutes a slow-fast system:
${ \delta F }$ evolves on a (fast) timescale of order ${ \tau \!\simeq\! 1/N }$,
while ${ F_{N} }$ evolves on a (slow) timescale of order ${ \tau \!\simeq\! 1 }$.

It is essential to note that Eqs.~\eqref{Klim_FastSlow_LDP}
are deterministic, but their initial conditions are random.
In the following, we assume that the dynamics of the fast process
is mixing, i.e.\ it forgets about initial conditions rapidly enough~\citep{Kifer2004}.
Though we cannot rigorously prove this mixing hypothesis
for the present system,
this seems like a natural assumption.

Benefiting from this separation of timescales, let us introduce a discrete timestep,
${ \Delta \tau }$, satisfying
\begin{equation}
1 / N \ll \Delta \tau \ll 1 .
\label{choice_Deltatau}
\end{equation}
We subsequently introduce the random variable, $G_{N}$,
associated with a finite-difference rate of change during a time ${ \Delta \tau }$.
It reads
\begin{equation}
G_{N} (\tau) = \frac{F_{N} (\tau \!+\! \Delta \tau) - F_{N} (\tau)}{\Delta \tau} .
\label{def_GN}
\end{equation}
The probability of a given time evolution, ${ \{ F_{N} (\tau) \}_{0 \leq \tau \leq T} }$,
can then be estimated
through the product of conditional probabilities
\begin{align}
\bbP \big[ {} &  \big\{ F_{N} (\tau) \big\}_{0 \leq \tau \leq T} \!=\! \big\{ F (\tau) \big\}_{0 \leq \tau \leq T} \big]  \;\; \underset{\mathclap{N \to + \infty}}{\asymp}
\label{calc_path}
\\
{} & \prod_{n} \bbP \big[ G_{N} ( n \Delta \tau) \!=\! \dot{F} (n \Delta \tau) \,\big|\, F_{N} (n \Delta \tau) \!=\! F (n \Delta \tau) \big] ,
\nonumber
\end{align}
with ${ \dot{F} \!=\! \p_{\tau} F }$.
Importantly, to obtain this expression,
we used a Markovian decomposition of the path
of evolution, assuming that increments of $F_{N}$
over timescales of order ${ \Delta \tau \!\gg\! 1/N }$,
are independent from one another.

Now, let us compute the probability of a given increment,
${ \bbP (G_{N} \!=\! \dot{F} \,|\, F_{N} \!=\! F) }$.
We use Eq.~\eqref{LDP_Gartner} with ${ N \Delta \tau \!\gg\! 1 }$
as the large deviation rate. We obtain
\begin{equation}
\bbP (G_{N} \!=\! \dot{F} \,|\, F_{N} \!=\! F) \;\;\;\underset{\mathclap{N \to + \infty}}{\asymp} \;\;\; \re^{- N \Delta \tau I [F , \dot{F}]} ,
\label{calc_Prod_G}
\end{equation}
where the dependence w.r.t.\ $F$ emphasises
that $F$ can be taken as constant on a time interval of duration ${ \Delta \tau }$.
In Eq.~\eqref{calc_Prod_G}, the large deviation function
follows from Eq.~\eqref{Legendre_2} and reads
\begin{equation}
I [F , \dot{F}] = \sup\limits_{P} \big\{ \dot{F} P \!-\! \mH[F , P] \big\} .
\label{calc_I_G_LDP}
\end{equation}
Here, ${ \mH [F,P] }$,
is the large deviation Hamiltonian
It follows from Eq.~\eqref{calc_Gartner_II},
with the large deviation rate, ${ \Delta t \!=\! N \Delta \tau }$,
and reads
\begin{equation}
\mH [F , P] = \frac{1}{\Delta t} \ln \big[ \big\langle \re^{\Delta t G_{N} P } \big\rangle_{\!F} \big] ,
\label{calc_H_G}
\end{equation}
where the average is performed over the fast process
from Eq.~\eqref{Klim_Fast_LDP} with ${ F_{N} \!=\! F }$.
Starting from the definition of Eq.~\eqref{def_GN},
we can write
\begin{equation}
G_{N} = \frac{1}{\Delta \tau} \!\! \int_{0}^{\Delta \tau} \!\! \rd \tau \, \p_{\tau} F_{N} = \frac{1}{\Delta t} \!\! \int_{0}^{\Delta t} \!\!\!\! \rd t \, \p_{\tau} F_{N} ,
\label{calc_GN}
\end{equation}
where, for simplicity, we started the time-integral in ${ \tau \!=\! 0 }$,
and ${ \p_{\tau} F_{N} }$ follows from Eq.~\eqref{Klim_Slow_LDP}.
Since ${ \Delta t \!\gg\! 1 }$ (Eq.~\ref{choice_Deltatau}),
we can approximate Eq.~\eqref{calc_H_G} with
\begin{equation}
\mH [F , P] = \!\lim\limits_{\Delta \to + \infty}\! \frac{1}{\Delta} \ln \bigg[ \bigg\langle \!\exp \!\bigg(\! P \!\! \int_{0}^{\Delta} \!\!\!\! \rd t \, \p_{\tau} F_{N} \!\bigg) \!\bigg\rangle_{\!F} \bigg] .
\label{final_H_LDP}
\end{equation}
The final step of the computation is to use the assumption
${ \Delta \tau \!\ll\! 1 }$ (Eq.~\ref{choice_Deltatau})
so that the product of probabilities in Eq.~\eqref{calc_path}
can be replaced by the exponential of an integral.
More precisely, using Eq.~\eqref{calc_Prod_G}, we obtain
\begin{align}
\bbP \big[ \big\{ F_{N} (\tau) {} & \big\}_{0 \leq \tau \leq T} \!=\! \big\{ F (\tau) \big\}_{0 \leq \tau \leq T} \big]  \;\;\; \underset{\mathclap{N \to + \infty}}{\asymp}
\label{final_LDP}
\\
{} & \exp \!\bigg[\! - N \sup_{P} \!\! \int_{0}^{T} \!\!\!\! \rd \tau \bigg\{ \dot{F} P - \mH [F , P] \bigg\} \bigg] ,
\nonumber
\end{align}
where ${ P \!=\! P (\tau) }$ is now a field w.r.t.\ the slow time $\tau$.
Naturally, Eqs.~\eqref{final_LDP} and~\eqref{final_H_LDP}
bear a lot of similarities with the generic results
from Eqs.~\eqref{large_deviations_generic} and~\eqref{def_mH_generic}.
To fully recover these expressions,
it only remains to
(i) add the additional dependence of ${ F_{N} }$
w.r.t.\ $\bJ$, which also transmits to ${ P \!=\! P (\bJ , \tau) }$;
(ii) account correctly for the various normalisation prefactors.
Let us conclude by pointing out that the present heuristic proof
does not check the differentiability of the large derivation Hamiltonian, ${ \mH [F , P] }$,
and hence the possible non-convexity
of the large deviation function, ${ I [F] }$.

\section{Computing the cumulants}
\label{app:Comp_Cum}

In this Appendix, we compute the first cumulants
of Eq.~\eqref{def_mH_generic}.
For a random variable $X$,
we recall that
\begin{equation}
\ln \big[ \big\langle \re^{X} \big\rangle \big] \simeq  \langle X \rangle + \tfrac{1}{2} \big( \langle X^{2} \rangle - \langle X \rangle^{2} \big) + ...
\label{cumulant_expansion}
\end{equation}
Following the convention from Eq.~\eqref{def_Fourier},
we note that Eq.~\eqref{Klim_Slow}
can be written as
\begin{equation}
\frac{\p F_{N} (\bJ)}{\p \tau} = - \ri \sum_{\bk} \bk \!\cdot\! \frac{\p }{\p \bJ} \bigg[ \delta F_{\bk} (\bJ) \, \delta \Phi_{- \bk} (\bJ) \bigg] .
\label{Fourier_Klim_Slow}
\end{equation}
Hence, we will apply Eq.~\eqref{cumulant_expansion}
using ${ X \!\propto\! \delta F \, \delta \Phi }$.

\subsection{First cumulant}
\label{app:Cum_1}

Owing to Eq.~\eqref{Fourier_Klim_Slow},
the first cumulant of Eq.~\eqref{def_mH_generic}
can be written as
\begin{equation}
\mH^{(1)} [F , P] = \sum_{\bk} \!\! \int \!\! \rd \bJ \, \bk \!\cdot\! \frac{\p P}{\p \bJ} \, C^{(1)}_{\bk} (\bJ) ,
\label{rewrite_H1}
\end{equation}
with the correlation function
\begin{equation}
C^{(1)}_{\bk} (\bJ) = \!\! \lim\limits_{\Delta \to + \infty} \! \ri \!\! \int_{0}^{\Delta} \!\!\!\! \rd t \, \big\langle \delta F_{\bk} (\bJ , t) \, \delta \Phi_{- \bk} (\bJ , t) \big\rangle_{\!F} .
\label{def_C1}
\end{equation}

The next step of the calculation is to inject
the time-dependent expression of the \DF\ and potential fluctuations,
from Eqs.~\eqref{time_deltaPhi} and~\eqref{time_dF}.
These fluctuations are expressed as a function of
the initial conditions, ${ \delta F_{\bk} (\bJ , 0) }$.
We assume that the initial fluctuations stem
from some uncorrelated Poisson shot noise,
hence making them Gaussian.
Following eq.~{(41)} of~\cite{Chavanis2012},
and paying attention to the prefactor ${ 1/\sqrt{N} }$
in Eq.~\eqref{decomposition_QL},
we have
\begin{equation}
\!\! \big\langle\! \delta F_{\bk} (\bJ , 0) \delta F_{\bkp} (\bJp , 0) \!\big\rangle_{\!F} \!=\! \Mtot \frac{\delta_{\bk, - \bkp}}{(2\pi)^{d}} \deltaD (\bJ \!-\! \bJp) F (\bJ) .
\label{Poisson_init}
\end{equation}

Using this statistics, Eq.~\eqref{def_C1} becomes
\begin{align}
{} & C^{(1)}_{\bk} (\bJ) = \ri \Mtot F (\bJ) \psi_{\bk\bk}^{\rd *} (\bJ , \bJ , \bk \!\cdot\! \bO)
\label{calc_same_correl}
\\
+ {} & \ri (2 \pi)^{d} \Mtot \bk \!\cdot\! \frac{\p F}{\p \bJ} \! \sum_{\bkp} \!\! \int \!\! \rd \bJp F (\bJp) \frac{|\psid_{\bk\bkp} (\bJ , \bJp , \bkp \!\cdot\! \bOp)|^{2}}{\bk \!\cdot\! \bO \!-\! \bkp \!\cdot\! \bOp \!-\! \ri \gamma} ,
\nonumber
\end{align}
where we used the symmetry
\begin{equation}
\psid_{-\bk-\bkp} (\bJ , \bJp , - \omegaR) = \psi^{\rd *}_{\bk\bkp} (\bJ , \bJp , \omegaR) ,
\label{symmetry_psid}
\end{equation}
for ${ \omegaR \!\in\! \mathbb{R} }$.

At this stage, following the truncation from Eq.~\eqref{exp_Landau},
we must compute Eq.~\eqref{calc_same_correl}
at order ${ o (G^{2}) }$.
We can generically expand the dressed coupling coefficient as
\begin{equation}
\psid_{\bk\bkp} (\bJ , \bJp , \omega) \!\simeq\! \psi_{\bk\bkp} (\bJ , \bJp) \!+\! \psi^{(2)}_{\bk\bkp} (\bJ , \bJp , \omega) \!+\! o (G^{2}) ,
\label{expansion_psid}
\end{equation}
with the scaling ${ \psi^{(2)}_{\bk\bkp} \!\propto\! G^{2} }$, for ${ G \!\to\! 0 }$.
Relying on the self-consistent relation from Eq.~\eqref{self_psid},
we get
\begin{equation}
\psi^{(2)}_{\bk\bk} \!(\bJ , \bJ , \bk \!\cdot\! \bO) \!=\! - (2 \pi)^{d} \!\sum_{\bkp}\! \!\! \int \!\! \rd \bJp \frac{|\psi_{\bk\bkp} (\bJ , \bJp)|^{2} \, \bkp \!\cdot\! \p F / \p \bJp}{\bk \!\cdot\! \bO \!-\! \bkp \!\cdot\! \bOp \!+\! \ri \gamma} ,
\label{calc_psi2}
\end{equation}
where we used the symmetry from Eq.~\eqref{sym_psi_order}
and the prescription from Eq.~\eqref{regularisation_denominator}.

At order ${ o (G^{2}) }$, Eq.~\eqref{calc_same_correl} then becomes
\begin{align}
C_{\bk}^{(1)} (\bJ) =  {} & \ri \Mtot F (\bJ) \psi_{\bk\bk}^{*} (\bJ , \bJ)
\nonumber
\\
+ {} & \ri (2 \pi)^{d} \Mtot \sum_{\bk} \!\! \int \!\! \rd \bJp \, \frac{|\psi_{\bk\bkp} (\bJ , \bJp)|^{2}}{\bk \!\cdot\! \bO \!-\! \bkp \!\cdot\! \bOp \!-\! \ri \gamma}
\nonumber
\\
\times {} & \bigg\{ \bk \!\cdot\! \frac{\p F}{\p \bJ} F(\bJp) \!-\! \bkp \!\cdot\! \frac{\p F}{\p \bJp} F (\bJ) \bigg\} ,
\label{calc_C1_next}
\end{align}
Performing the symmetrisation ${ \bk \!\to\! - \bk }$ in Eq.~\eqref{rewrite_H1},
and using the symmetry from Eq.~\eqref{symmetry_psi},
we find that the term in ${ \psi_{\bk\bk}^{*} (\bJ , \bJ) }$ in Eq.~\eqref{calc_C1_next}
does not contribute to $\mH^{(1)}$.
The final step of the calculation is to expand the resonant denominator
using Plemelj formula
\begin{equation}
\frac{1}{\omegaR \!-\! \ri \gamma} = \mP \bigg( \frac{1}{\omegaR} \bigg) + \ri \pi \deltaD (\omegaR) ,
\label{Plemelj}
\end{equation}
with $\mP$ Cauchy principal value.
Performing the symmetrisation ${ (\bk,\bkp) \!\to\! (- \bk,-\bkp) }$ in Eq.~\eqref{rewrite_H1}
and~\eqref{calc_C1_next},
and using once again the symmetries from Eq.~\eqref{symmetry_psi},
we find that the principal value does not contribute to $\mH^{(1)}$.
Ultimately, at order ${ o (G^{2}) }$, we are left with
\begin{align}
\mH^{(1)} [F , P] = - {} & \sum_{\bk , \bkp} \! \int \!\! \rd \bJ \rd \bJp \, \bk \!\cdot\! \frac{\p P}{\p \bJ} \, B_{\bk\bkp} (\bJ , \bJp)
\nonumber
\\
{} & \times \, \bigg\{ \bk \!\cdot\! \frac{\p F}{\p \bJ} F (\bJp) \!-\! \bkp \!\cdot\! \frac{\p F}{\p \bJp} F (\bJ) \bigg\} ,
\label{final_mH_1}
\end{align}
hence recovering Eq.~\eqref{mH_1}.

\subsection{Second cumulant}
\label{app:Cum_2}

Let us now compute the second cumulant of Eq.~\eqref{def_mH_generic}.
In that case, following Eq.~\eqref{cumulant_expansion},
we must compute averages of order ${ X^{2} }$, with ${ X \!\propto\! \delta F \, \delta \Phi }$.
Since we are computing expressions at order ${ o (G^{2}) }$,
we follow the scalings from Eq.~\eqref{scaling_deltaPhi_deltaF}
to simplify the expressions of ${ \delta \Phi (t) }$ and ${ \delta F (t) }$
to be used in this computation.
More precisely, at order ${ o (G^{2}) }$,
we may replace Eq~\eqref{time_deltaPhi} with
\begin{align}
\delta \Phi_{\bk} (\bJ , t) = (2 \pi)^{d} \sum_{\bkp} \!\! \int {} & \!\! \rd \bJp \, \re^{- \ri \bkp \cdot \bOp t} \delta F_{\bkp} (\bJp , 0)
\nonumber
\\
\times {} & \, \psi_{\bk\bkp} (\bJ , \bJp) ,
\label{deltaPhi_H2}
\end{align}
and Eq.~\eqref{time_dF} with
\begin{equation}
\delta F_{\bk} (\bJ , t) = \re^{- \ri \bk \cdot \bO t} \, \delta F_{\bk} (\bJ, 0 ) .
\label{deltaF_H2}
\end{equation}
This is a key step to simplify
the upcoming calculations.

The second cumulant of Eq.~\eqref{def_mH_generic} then reads
\begin{equation}
\mH^{(2)} [F , P] \!=\! \sum_{\bk,\bkp} \! \int \!\! \rd \bJ \rd \bJp \, \bk \!\cdot\! \frac{\p P}{\p \bJ} \, \bkp \!\cdot\! \frac{\p P}{\p \bJp} \, C^{(2)}_{\bk\bkp} (\bJ , \bJp) ,
\label{calc_mH2}
\end{equation}
where we introduced the correlation function
\begin{align}
{} & C^{(2)}_{\bk\bkp} (\bJ , \bJp) = - \lim\limits_{\Delta \to + \infty} \frac{1}{2} \frac{(2 \pi)^{d}}{\Delta\Mtot} \!\! \int_{0}^{\Delta} \!\!\!\! \rd t \!\! \int_{0}^{\Delta} \!\!\!\! \rd \tp 
\label{def_C2}
\\
{} & \times \big\langle\!\big\langle \big[ \delta F_{\bk} (\bJ , t) \, \delta \Phi_{- \bk} (\bJ , t) \big] \big[ \delta F_{\bkp} (\bJp , \tp) \, \delta \Phi_{- \bkp} (\bJp , \tp) \big] \big\rangle\!\big\rangle_{\!F} ,
\nonumber
\end{align}
with the notation ${ \langle\!\langle X Y \rangle\!\rangle_{F} \!=\! \langle X Y \rangle_{F} \!-\! \langle X \rangle_{F} \langle Y \rangle_{F} }$.

Injecting the dependence from Eqs.~\eqref{deltaPhi_H2} and~\eqref{deltaF_H2},
we can rewrite Eq.~\eqref{def_C2} as
\begin{align}
C^{(2)}_{\bk\bkp} {} & (\bJ , \bJp) = - \!\! \lim\limits_{\Delta \to + \infty} \! \frac{1}{2}\frac{(2 \pi)^{3d}}{\Delta \Mtot} \!\! \int_{0}^{\Delta} \!\!\!\! \rd t \!\! \int_{0}^{\Delta} \!\!\!\! \rd \tp \!\! \sum_{\bk_{1} , \bkp_{1}} \!\! \int \!\! \rd \bJ_{1} \rd \bJp_{1}
\nonumber
\\
\times {} & \re^{- \ri \bk \cdot \bO t} \re^{- \ri \bk_{1} \cdot \bO_{1} t} \psi_{-\bk\bk_{1}} (\bJ , \bJ_{1})
\label{calc_C2}
\\
\times {} & \re^{- \ri \bkp \cdot \bOp \tp} \re^{- \ri \bkp_{1} \cdot \bOp_{1} \tp} \psi_{-\bkp\bkp_{1}} (\bJp , \bJp_{1})
\nonumber
\\
\times {} & \big\langle\!\big\langle \big[ \delta F_{\bk} (\bJ , 0) \, \delta F_{\bk_{1}} (\bJ_{1} , 0) \big]  \big[ \delta F_{\bkp} (\bJp , 0) \, \delta F_{\bkp_{1}} (\bJp_{1} , 0) \big] \big\rangle\!\big\rangle_{\!F}
\nonumber
\end{align}

Computing the average from Eq.~\eqref{def_C2}
requires the computation of the four-point correlation of ${ \delta F_{\bk} (\bJ , 0) }$.
We assume that the initial fluctuations are Gaussian so that
writing, ${ \delta F (1) \!=\! \delta F_{\bk_{1}} (\bJ_{1} , 0) }$,
we have
\begin{align}
\big\langle\!\big\langle \big[ \delta F (1) \, \delta F (2) \big] \big[ {} & \delta F (3) \, \delta F (4) \big] \big\rangle\!\big\rangle_{\!F} =
\label{4point}
\\
+ {} & \, \big\langle \delta F(1) \, \delta F (3) \big\rangle_{\!F} \, \big\langle \delta F (2) \delta F (4) \big\rangle_{\!F}
\nonumber
\\
+ {} & \, \big\langle \delta F(1) \, \delta F (4) \big\rangle_{\!F} \, \big\langle \delta F (2) \delta F (3) \big\rangle_{\!F} .
\nonumber
\end{align}

Following Eq.~\eqref{Poisson_init}, we can then rewrite Eq.~\eqref{calc_C2} as
\begin{align}
{} & C^{(2)}_{\bk\bkp} (\bJ , \bJp) = - \!\! \lim\limits_{\Delta \to + \infty} \! \frac{(2 \pi)^{d} \Mtot}{2 \Delta} \!\! \int_{0}^{\Delta} \!\!\!\! \rd t \!\! \int_{0}^{\Delta} \!\!\!\! \rd \tp 
\nonumber
\\
\times {} & \bigg\{ \re^{- \ri (\bk \cdot \bO - \bkp \cdot \bOp) (t - \tp)} \, | \psi_{\bk\bkp} (\bJ , \bJp) |^{2} \, F (\bJ) \, F (\bJp)
\nonumber
\\
+ {} & \, \delta_{\bk, - \bkp} \deltaD (\bJ \!-\! \bJp) \, F (\bJ)
\label{calc_C2_next}
\\
\times {} & \sum_{\bkpp} \!\! \int \!\! \rd \bJpp \, \re^{- \ri (\bk \cdot \bO - \bkpp \cdot \bOpp) (t - \tp)} \, | \psi_{\bk\bkpp} (\bJ , \bJpp) |^{2} F (\bJpp)
 \bigg\} ,
 \nonumber
\end{align}
where we used the symmetries from Eq.~\eqref{symmetry_psi}.

We now use the identity
\begin{align}
\lim\limits_{\Delta \to + \infty} \frac{1}{\Delta} \!\! \int_{0}^{\Delta} \!\!\!\! \rd t \!\! \int_{0}^{\Delta} \!\!\!\! \rd \tp \, \re^{- \ri \omegaR (t - \tp)} {} & = \!\! \int_{- \infty}^{+ \infty} \!\!\!\! \rd t \, \re^{- \ri \omegaR t}
\nonumber
\\
{} & = 2 \pi \deltaD (\omegaR) ,
\label{id_time_integral}
\end{align}
and Eq.~\eqref{calc_C2_next} becomes
\begin{align}
C^{(2)}_{\bk\bkp} (\bJ , {} & \bJp) = - B_{\bk\bkp} (\bJ , \bJp) F (\bJ) F (\bJp)
\label{calc_C2_nextnext}
\\
- {} & \delta_{\bk, -\bkp} \deltaD (\bJ \!-\! \bJp) \! \sum_{\bkpp} \!\! \int \!\! \rd \bJpp B_{\bk\bkpp} (\bJ , \bJpp)  F (\bJ) F (\bJpp) .
\nonumber
\end{align}
where we used the definition of ${ B_{\bk\bkp} (\bJ , \bJp) }$
from Eq.~\eqref{def_B}.

The last step of the calculation is to inject Eq.~\eqref{calc_C2_nextnext}
into Eq.~\eqref{calc_mH2}. We obtain
\begin{align}
\mH^{(2)} [F , P] = {} & \sum_{\bk , \bkp} \! \int \!\! \rd \bJ \rd \bJp \, B_{\bk\bkp} (\bJ , \bJp) \, \bk \!\cdot\! \frac{\p P}{\p \bJ}
\nonumber
\\
{} & \times \,  \bigg[ \bk \!\cdot\! \frac{\p P}{\p \bJ} - \bkp \!\cdot\! \frac{\p P}{\p \bJp} \bigg] \, F (\bJ) \, F (\bJp) ,
\label{final_calc_mH2}
\end{align}
hence recovering Eq.~\eqref{mH_2}.

\subsection{High-order cumulants}
\label{app:Cum_next}

Following Eq.~\eqref{cumulant_expansion},
cumulants of order higher than two involve averages
of order ${ X^{k} }$, with ${ k \!\geq\! 3 }$
and ${ X \!\propto\! \delta F \delta \Phi }$.
Following Eq.~\eqref{scaling_deltaPhi_deltaF},
we have ${ \delta \Phi \!=\! \mO (G) }$ for ${ G \!\to\! 0 }$.
Therefore, one has ${ X^{k} \!=\! \mO (G^{k}) \!=\! o (G^{2}) }$,
for ${ k \!\geq\! 3 }$.
As a conclusion,
in the dynamically hot limit,
cumulants of order higher than two do not contribute
to the large deviation Hamiltonian from Eq.~\eqref{mH_Landau}.

\section{Properties}
\label{app:Properties}

In this Appendix, we briefly justify the various properties
of the large deviation Hamiltonian from Eq.~\eqref{mH_Landau}.

\subsection{Most probable path}
\label{app:MostProbable}

As required by Eq.~\eqref{most_probable},
we need to compute the functional gradient
${ \delta \mH / \delta P (\bJ) }$.
To do so, we rely on the fundamental identity
\begin{equation}
\frac{\delta P (\bJp)}{\delta P (\bJ)} = \deltaD (\bJ \!-\! \bJp) .
\label{fundamental_identity_grad}
\end{equation}
Since ${ \mH^{(2)} }$ is quadratic in ${ P (\bJ) }$
(Eq.~\ref{mH_2}), one has ${ \p \mH^{(2)} / \delta P (\bJ) \!=\! 0 }$
in ${ P \!=\! 0 }$. As a consequence, only ${ \mH^{(1)} }$
contributes to the most probable path.
Integrating by parts w.r.t.\ ${ \rd \bJ }$
the term ${ \bk \!\cdot\! \p P / \p \bJ }$ in Eq.~\eqref{mH_1}
readily recovers the inhomogeneous Landau
equation~\eqref{exp_Landau}.

\subsection{Conservation laws}
\label{app:Conservation}

\textbf{Mass conservation}. Following Eq.~\eqref{def_C_mass},
we have ${ \delta M / \delta F (\bJ) \!=\! 1 }$.
We note from Eqs.~\eqref{mH_1} and~\eqref{mH_2}
that the large deviation Hamiltonian
only depends on derivatives of the conjugate field, ${ P (\bJ) }$.
To check for mass conservation, we compute terms of the form
\begin{align}
\!\! \int \!\! \rd \bJ \, \frac{\delta M}{\delta F (\bJ)} \frac{\delta \mH}{\delta P (\bJ)} {} & = \!\! \int \!\! \rd \bJ \rd \bJp \, \frac{\delta }{\delta P (\bJ)} \bigg[ \frac{\p P}{\p \bJp} \bigg] \times ...
\nonumber
\\
{} & = \!\! \int \!\! \rd \bJ \rd \bJp \, \frac{\p }{\p \bJp} \bigg[ \deltaD (\bJ \!-\! \bJp) \bigg] \times ...
\nonumber
\\
{} & = - \!\! \int \!\! \rd \bJ \, \frac{\p }{\p \bJ} \bigg[ ... \bigg]
\nonumber
\\
{} & = 0 ,
\label{shape_calc_M}
\end{align}
This ensures consistence w.r.t.\ mass conservation.

\textbf{Energy conservation}. Following Eq.~\eqref{def_C_energy},
we have ${ \delta E / \delta F (\bJ) \!=\! H (\bJ) }$.
Following some integration by parts and manipulations,
we get from Eq.~\eqref{mH_1}
\begin{align}
\!\! \int \!\! \rd \bJ \, \frac{\delta E}{\delta F(\bJ)} \frac{\delta \mH^{(1)}}{\delta P (\bJ)} {} & =  - \sum_{\bk,\bkp} \! \int \!\! \rd \bJ \rd \bJp \, B_{\bk\bkp} (\bJ , \bJp) \, \bk \!\cdot\! \bO
\nonumber
\\
\times {} & \bigg[ \bk \!\cdot\! \frac{\p F}{\p \bJ} F (\bJp) \!-\! \bkp \!\cdot\! \frac{\p F}{\p \bJp} F (\bJ) \bigg] . 
\label{calc_E_H1}
\end{align}
Similarly, we get from Eq.~\eqref{mH_2}
\begin{align}
\!\! \int \!\! \rd \bJ \, \frac{\delta E}{\delta F(\bJ)} \frac{\delta \mH^{(2)}}{\delta P(\bJ)} {} & = \sum_{\bk,\bkp} \! \int \!\! \rd \bJ \rd \bJp \, B_{\bk\bkp} (\bJ , \bJp) \, \bk \!\cdot\! \bO
\nonumber
\\
\times {} & 2 F (\bJ) F (\bJp) \bigg[ \bk \!\cdot\! \frac{\p P}{\p \bJ} \!-\! \bkp \!\cdot\! \frac{\p P}{\p \bJp} \bigg] .
\label{calc_E_H2}
\end{align}

We now perform the symmetrisation ${ (\bk , \bJ) \!\leftrightarrow\! (\bkp, \bJp) }$
in Eqs.~\eqref{calc_E_H1} and~\eqref{calc_E_H2}
and rely on the symmetry from Eq.~\eqref{symmetry_psi}.
Both equations then involve
\begin{equation}
B_{\bk\bkp} (\bJ , \bJp) \big[ \bk \!\cdot\! \bO \!-\! \bkp \!\cdot\! \bOp \big] = 0 ,
\label{symmetry_B_energy}
\end{equation}
which vanishes owing to the resonance condition from Eq.~\eqref{def_B}.
As a conclusion, we therefore have
\begin{equation}
\!\! \int \!\! \rd \bJ \, \frac{\delta E}{\delta F(\bJ)} \frac{\p \mH}{\p P (\bJ)} = 0 .
\label{final_E}
\end{equation}
This ensures consistence w.r.t.\ energy conservation.

\subsection{Hamilton--Jacobi equation}
\label{app:HamiltonJacobi}

Following Eq.~\eqref{def_S},
we have
\begin{equation}
\frac{\delta S}{\delta F (\bJ)} = - \ln [F (\bJ)] + \cst.
\label{der_S}
\end{equation}
Since ${ \mH [F , P] }$ only depends on gradients of $P$
(see Appendix~\ref{app:Conservation}),
the constant term in Eq.~\eqref{der_S} does not contribute
to ${ \mH [F , - \delta S / \delta F] }$.
Following Eq.~\eqref{mH_1}, the first cumulant contributes
\begin{align}
\mH^{(1)} [F , - \delta S / \delta F] \!= - {} & \! \sum_{\bk, \bkp} \! \int \!\! \rd \bJ \rd \bJp B_{\bk\bkp} \!(\bJ , \bJp) \bk \!\cdot\! \frac{\p \ln [F (\bJ)]}{\p \bJ}
\nonumber
\\
\times {} & \bigg[ \bk \!\cdot\! \frac{\p F}{\p \bJ} F (\bJp) \!-\! \bkp \!\cdot\! \frac{\p F}{\p \bJ} F (\bJ) \bigg] .
\label{calc_S_1}
\end{align}
Noting that ${ B_{\bkp\bk} (\bJp , \bJ) \!=\! B_{\bk\bkp} (\bJ , \bJp) }$
(Eq.~\ref{def_B}), we symmetrise Eq.~\eqref{calc_S_1} with ${ (\bk , \bJ) \!\leftrightarrow\! (\bkp, \bJp) }$ to get
\begin{align}
\mH^{(1)} [F , - \delta S / \delta F] {} & = - \sum_{\bk , \bkp} \! \int \!\! \rd \bJ \rd \bJp \, \frac{B_{\bk\bkp} (\bJ , \bJp)}{F (\bJ) F (\bJp)}
\nonumber
\\
\times {} & \bigg[ \bk \!\cdot\! \frac{\p F}{\p \bJ} F (\bJp) \!-\! \bkp \!\cdot\! \frac{\p F}{\p \bJp} F(\bJ) \bigg]^{2} .
\label{calc_S_1_next}
\end{align}

Following Eq.~\eqref{mH_2}, the second cumulant contributes
\begin{align}
\mH^{(2)} {} & [F , - \delta S / \delta F] = \sum_{\bk , \bkp} \! \int \!\! \rd \bJ \rd \bJp \, B_{\bk\bkp} (\bJ , \bJp) \, \bk \!\cdot\! \frac{\p \ln [F(\bJ)]}{\p \bJ}
\nonumber
\\
\times {} & \bigg[ \bk \!\cdot\! \frac{\p \ln [F (\bJ)]}{\p \bJ} \!-\! \bkp \!\cdot\! \frac{\p \ln [F(\bJp)]}{\p \bJp}  \bigg] F(\bJ) F (\bJp) .
\label{calc_S_2}
\end{align}
Performing the same symmetrisation as in Eq.~\eqref{calc_S_1_next},
we are left with
\begin{align}
\mH^{(2)} [F , - \delta S / \delta F] {} & = \sum_{\bk , \bkp} \! \int \!\! \rd \bJ \rd \bJp \, \frac{B_{\bk\bkp} (\bJ , \bJp)}{F (\bJ) F (\bJp)}
\nonumber
\\
\times {} & \bigg[ \bk \!\cdot\! \frac{\p F}{\p \bJ} F (\bJp) \!-\! \bkp \!\cdot\! \frac{\p F}{\p \bJp} F (\bJ) \bigg]^{2} .
\label{calc_S_2_next}
\end{align}
Combining Eqs.~\eqref{calc_S_1_next} and~\eqref{calc_S_2_next}, 
we ultimately obtain the expected relation
\begin{align}
\mH [F , - \delta S / \delta F] {} & = \mH^{(1)} [F , - \delta S / \delta F] \!+\! \mH^{(2)} [F , - \delta S / \delta F]
\nonumber
\\
{} & = 0 .
\label{calc_S_final}
\end{align}

\subsection{Time reversal symmetry}
\label{app:TimeReversal}

Given that ${ \mH^{(1)} [F,P] }$ (resp.\ ${ \mH^{(2)} [F, P] }$)
is linear (resp.\ quadratic) w.r.t.\ $P$, we immediately have
\begin{equation}
\mH [F , - P] = - \mH^{(1)} [F , P] + \mH^{(2)} [F , P] .
\label{neg_H}
\end{equation}
Similarly, by linearity we have
\begin{equation}
\mH^{(1)} \![ F , P \!-\! \delta S / \delta F ] \!=\! \mH^{(1)} \![ F , P ] \!+\! \mH^{(1)} \![ F, - \delta S / \delta F ] . 
\label{neg_H1_diff}
\end{equation}
As for the second cumulant, it reads
\begin{align}
 \mH^{(2)} [ F {} & , P \!-\! \delta S / \delta F ] \!=\!  \mH^{(2)} [ F , P ] \!+\! \mH^{(2)} [ F , - \delta S / \delta F ]
\nonumber
\\
+ {} & \tmH^{(2)} [ F , P , - \delta S / \delta F ] \!+\! \tmH^{(2)} [ F , - \delta S / \delta F , P ] ,
\label{neg_H2_diff}
\end{align}
where, following Eq.~\eqref{mH_2}, we introduced
\begin{align}
\tmH^{(2)} [F , P , Q] = \!\! \sum_{\bk , \bkp} {} & \! \int \!\! \rd \bJ \rd \bJp \, B_{\bk\bkp} (\bJ , \bJp) \, \bk \!\cdot\! \frac{\p P}{\p \bJ}
\nonumber
\\
\times {} & \bigg[ \bk \!\cdot\! \frac{\p Q}{\p \bJ} \!-\! \bkp \!\cdot\! \frac{\p Q}{\p \bJp} \bigg] \, F (\bJ ) \, F (\bJp) .
\label{def_tH2}
\end{align}

With the same symmetrisation as in Eq.~\eqref{calc_S_1_next},
one gets
\begin{subequations}
\begin{align}
\tmH^{(2)} [ F , P , - \delta S / \delta F] = - \mH^{(1)} [F , P] , 
\label{symm_tmH_1}
\\
\tmH^{(2)} [F , - \delta S / \delta F , P] = - \mH^{(1)} [F , P] . 
\end{align}
\label{symm_tmH}\end{subequations}
Recalling the result from Eq.~\eqref{calc_S_final},
all the previous relations lead to the needed result,
namely
\begin{equation}
\mH [F , - P] = \mH [F , P \!-\! \delta S / \delta F] . 
\label{calc_sym_time_final}
\end{equation}

\subsection{Gradient structure}
\label{app:Gradient}

The expression of ${ Q[F] }$ in Eq.~\eqref{def_Q}
follows from integration by parts of Eq.~\eqref{mH_2}.
We now compute the r.h.s.\ of Eq.~\eqref{Grad_Landau}.
Following Eq.~\eqref{der_S}, we face the term
\begin{align}
\!\! \int \!\! \rd \bJp \, Q [F] (\bJ , \bJp) \, \frac{\p S [F]}{\p F (\bJp)} \!=\! {} & \!\! \int \!\! \rd \bJp \, \bk \!\cdot\! \frac{\p }{\p \bJ} \bigg\{ \frac{\p }{\p \bJp} \!\cdot\! \bA (\bJ , \bJp) \bigg\}
\nonumber
\\
\times {} & \bigg\{ - \ln [F (\bJp)] + \cst \bigg\} ,                                                                                                                                                     
\label{calc_Grad}
\end{align}
where ${ \bA (\bJ , \bJp) }$ follows from Eq.~\eqref{def_Q}.
Discarding boundary terms,
we find that the constant terms do not contribute.
Integrating by parts w.r.t.\ ${ \rd \bJp }$,
we get
\begin{equation}
\eqref{calc_Grad} = \bk \!\cdot\! \frac{\p }{\p \bJ} \bigg\{ \!\! \int \!\! \rd \bJp \, \frac{1}{F (\bJp)} \, \bA (\bJ , \bJp) \!\cdot\! \frac{\p F}{\p \bJp} \bigg\} .
\label{calc_Grad_next}
\end{equation}
Injecting ${ \bA (\bJ , \bJp) }$ from Eq.~\eqref{def_Q},
one recovers Eq.~\eqref{Grad_Landau}.

\subsection{Stochastic Landau equation}
\label{app:Stochastic}

As defined in Eq.~\eqref{def_zeta},
the correlation function of the stochastic noise,
${ \zeta[F] (\bJ , \tau) }$,
involves Dirac deltas.
This makes the sampling of effective realisations challenging.
In addition, although guaranteed by Eq.~\eqref{cons_C},
it is not strikingly obvious that Eq.~\eqref{SPDE},
indeed, complies with the system's conservation laws.
In this Appendix, we tackle these two issues
and devise a ``diagonal'' rewriting of ${ \zeta[F] (\bJ , \tau) }$:
it is sourced by a normal random Gaussian field
and makes the conservation laws obvious.

For any given resonance vector $\bk$,
we assume that we can perform the change of variables
${ \bJ \!\mapsto\! (\omega \!=\! \bk \!\cdot\! \bO , \bz) }$~\citep[see, e.g.\@, ][for an explicit example]{Fouvry+2022}
with a non-degenerate Jacobian,
${ \mJ \!=\! | \p (\omega , \bz) / \p \bJ | }$.
Physically, ${ \omega }$ is the resonance frequency,
and $\bz$ covers the sets of all orbits that resonate with it.
The existence of the present mapping is mandatory for the resonance condition,
${ \deltaD (\bk \!\cdot\! \bO \!-\! \bkp \!\cdot\! \bOp) }$,
to be generically well-posed in the Landau Eq.~\eqref{exp_Landau}.

Let us consider the following ansatz for the noise
\begin{align}
\zeta [F] (\bJ , \tau) {} & \!=\! \frac{\sqrt{\pi} \Mtot}{\sqrt{N}} \frac{\p }{\p \bJ} \!\cdot\! \bigg[ \sum_{\bk , \bkp} \bk \!\! \int \!\! \rd \bJp \, \big[ \mJ \mJp \big]^{1/2}
\nonumber
\\
\times {} & | \psi_{\bk\bkp} (\bJ , \bJp) | \big[ F (\bJ) F (\bJp) \big]^{1/2} \deltaD (\omega \!-\! \omegap) 
\nonumber
\\
\times {} & \big\{ \eta_{\bk\bkp} (\bz , \bzp ,\omega , \tau) \!-\! \eta_{\bkp\bk} (\bzp , \bz , \omegap , \tau) \big\} \bigg] ,
\label{diag_zeta}
\end{align}
where we used shortened notations
for the change of variables, ${ \bJ \leftrightarrow_{\bk} \!(\omega , \bz) }$
and its Jacobian $\mJ$,
and similarly for $\bJp$.
In Eq.~\eqref{diag_zeta}, we also introduced the normal Gaussian random field,
${ \eta_{\bk\bkp} (\bz , \bzp , \omega , \tau) }$, obeying
\begin{subequations}
\begin{align}
{} & \big\langle \eta_{\bk\bkp} (\bz, \bzp , \omega , \tau) \big\rangle = 0 ,
\label{def_eta_1}
\\
{} & \big\langle \eta_{\bk\bk_{1}} \!(\bz , \bz_{1} , \omega , \tau) \, \eta_{\bkp\bkp_{1}} \!(\bzp , \bzp_{1} , \omegap , \taup) \big\rangle \!=\! \delta_{\bk\bkp} \delta_{\bk_{1} \bkp_{1}}
\label{def_eta_2}
\\
{} & \;\;\;\;\;\;\; \times \deltaD (\bz \!-\! \bzp) \, \deltaD (\bz_{1} \!-\! \bzp_{1}) \, \deltaD (\omega \!-\! \omegap) \, \deltaD (\tau \!-\! \taup) .
\nonumber
\end{align}
\label{def_eta}\end{subequations}

Equation~\eqref{diag_zeta}
easily complies with the conservation laws
from Eqs.~\eqref{def_C}.
Indeed, Eq.~\eqref{diag_zeta} is the divergence
of a flux in action space, hence the total mass is conserved.
As for the total energy,
starting from Eq.~\eqref{def_C_energy},
up to prefactors,
one must compute
\begin{align}
\frac{\rd E}{\rd t} \propto {} & \sum_{\bk , \bkp} \! \int \!\! \rd \bJ \rd \bJp \, \omega \, \deltaD (\omega \!-\! \omegap) \, \big[ \mJ \mJp \big]^{1/2}
\nonumber
\\
\times {} & | \psi_{\bk\bkp} (\bJ , \bJp) | \, \big[ F (\bJ) F(\bJp) \big]^{1/2}
\nonumber
\\
\times {} & \big\{ \eta_{\bk\bkp} (\bz , \bzp , \omega , \tau) \!-\! \eta_{\bkp \bk} (\bzp , \bz , \omegap , \tau) \big\} .
\label{calc_dEdt_eta}
\end{align}
Performing the symmetrisation ${ (\bk , \bJ) \!\leftrightarrow\! (\bkp , \bJp) }$
leaves us with ${ (\omega \!-\! \omegap) \, \deltaD (\omega \!-\! \omegap) \!=\! 0 }$,
so that ${ \rd E / \rd t \!=\! 0 }$.

Let us now check Eqs.~\eqref{def_zeta}.
Since ${ \zeta[F] }$ is linear w.r.t.\ ${ \eta }$,
Eq.~\eqref{def_eta_1} naturally imposes
${ \langle \zeta[F] \rangle \!=\! 0 }$, as required
by Eq.~\eqref{def_zeta_1}.
As for the correlation function, we write
\begin{align}
{} & \big\langle \zeta[F] (\bJ , \tau) \, \zeta[F] (\bJp , \taup) \big\rangle \!=\! \pi m \Mtot \!\sum_{\mathclap{\substack{\bk , \bk_{1} \\ \bkp , \bkp_{1}}}} \!\! \int \!\! \rd \bJ_{1} \rd \bJp_{1} \bk \!\cdot\! \frac{\p }{\p \bJ} \bigg\{ 
\nonumber
\\
{} & \times \bkp \!\cdot\! \frac{\p }{\p \bJp} \bigg[ \big[ \mJ \mJ_{1} \mJp \! \mJp_{1} \big]^{1/2} |\psi_{\bk\bk_{1}} (\bJ , \bJ_{1})| |\psi_{\bkp\bkp_{1}} (\bJp , \bJp_{1})|
\nonumber
\\
{} & \times \deltaD (\omega \!-\! \omega_{1}) \deltaD (\omegap \!-\! \omegap_{1}) \big[ F (\bJ) F (\bJ_{1}) F(\bJp) F (\bJp_{1}) \big]^{1/2}
\nonumber
\\
{} & \times \big\langle \big\{ \eta_{\bk\bk_{1}} (\bz , \bz_{1} , \omega , \tau) \!-\! \eta_{\bk_{1} \bk} (\bz_{1} , \bz , \omega_{1} , \tau) \big\}
\nonumber
\\
{} & \times \big\{ \eta_{\bkp\bkp_{1}} (\bzp , \bzp_{1} , \omegap , \taup) \!-\! \eta_{\bkp_{1} \bkp} (\bzp_{1} , \bzp , \omegap_{1} , \taup) \big\} \big\rangle \bigg] \bigg\} ,
\label{calc_zeta_sq}
\end{align}
with
${ \bJ \leftrightarrow_{\bk} \!(\omega , \bz) }$
and similarly for $\bJp$, $\bJ_{1}$, $\bJp_{1}$.

In Eq.~\eqref{calc_zeta_sq}, to compute a given crossed term, say
${ \langle \eta_{\bk\bk_{1}} \eta_{\bkp\bkp_{1}} \rangle }$,
we use Eq.~\eqref{def_eta_2} and face a product
of three resonant Dirac deltas in frequencies.
We write it as
\begin{align}
\deltaD (\omega \!-\! \omega_{1}) \, {} & \deltaD (\omegap \!-\! \omegap_{1}) \, \deltaD (\omega \!-\! \omegap) = 
\label{prod_Dirac_1}
\\
{} & \deltaD(\omega \!-\! \omegap) \, \deltaD (\omega_{1} \!-\! \omegap_{1}) \, \deltaD (\omega \!-\! \omega_{1}) .
\nonumber
\end{align}
For the other crossed terms,
we pick the appropriate set
of differences of frequencies.
Finally,
we also use
\begin{equation}
\big[ \mJ \mJp \big]^{1/2} \delta_{\bk\bkp} \deltaD(\omega \!-\! \omegap) \deltaD (\bz \!-\! \bzp) \!=\! \delta_{\bk\bkp} \deltaD (\bJ \!-\! \bJp) ,
\label{volume_Dirac}
\end{equation}
and similar variations.
Following simple manipulations,
one ultimately recovers Eq.~\eqref{def_zeta_2}.

\end{document}